\documentclass[twocolumn,eqsecnum,showpacs,preprintnumbers,amsmath,amssymb]{revtex4}
\usepackage{latexsym} 
\usepackage{graphicx}
\usepackage{slashed}
\def\qed{\hbox{${\vcenter{\vbox{                          
   \hrule height 0.4pt\hbox{\vrule width 0.4pt height 6pt
   \kern5pt\vrule width 0.4pt}\hrule height 0.4pt}}}$}}
\def\shiftdown#1{#1\llap{\lower.04ex\hbox{#1}}}

\def\be{\begin{equation}}
\def\ee{\end{equation}}
\def\bea{\begin{eqnarray}}
\def\eea{\end{eqnarray}}
\begin{document}
\title{Hyperquarks and bosonic preon bound states}
\renewcommand{\thefootnote}{\fnsymbol{footnote}}
\author{Michael L. Schmid 
and Alfons J. Buchmann}
\affiliation{Institut f\"ur Theoretische Physik, Universit\"at T\"ubingen
         Auf der Morgenstelle 14, D-72076 T\"ubingen, Germany}
\email{alfons.buchmann@uni-tuebingen.de}
\email{micha.l.schmid@gmx.net} 

\begin{abstract}
In a model in which leptons, quarks, and the recently introduced hyperquarks 
are built up from two fundamental spin-$\frac{1}{2}$ preons, the standard model
weak gauge bosons emerge as preon bound states. In addition, the model predicts
a host of new composite gauge bosons, in particular those responsible for 
hyperquark and proton decay. Their presence entails a left-right symmetric 
extension of the standard model weak interactions and a scheme for a partial 
and grand unification of nongravitational interactions based on respectively
the effective gauge groups SU(6)$_P$ and SU(9)$_G$. This leads to a prediction 
of the Weinberg angle at low energies in good agreement with experiment.  
Furthermore, using evolution equations for the effective coupling strengths, 
we calculate the partial and grand unification scales, the hyperquark mass 
scale, as well as the mass and decay rate of the lightest hyperhadron. 
\end{abstract}

\smallskip

\maketitle

\section{Introduction}
\label{sec:introduction}
In a composite model in which leptons and quarks are bound states of two
fundamental, massless spin-$\frac{1}{2}$ preons, called $T$ and $V$,
interacting via color and hypercolor forces~\cite{har79,har80a,har81a},
a new class of fermionic bound states, called hyperquarks, 
has recently been introduced in order to satisfy a special case of the
't Hooft anomaly matching conditions~\cite{abms05}. 
In contrast to quarks, which carry zero hypercolor and open color,  
hyperquarks have zero color and open hypercolor. 
The matching of the anomalies on the preon and bound state levels that has been
achieved by introducing hyperquarks gives an answer to the question 
why there are exactly three fermionic generations.
At the same time it raises new questions, 
foremost whether there is any experimental evidence for hyperquarks.

Because hyperquarks are subject to confining forces one would expect 
hyperquark bound states, such as hypermesons and hyperbaryons to exist in 
nature. The nonobservation of these hyperhadrons 
with present accelerators indicates that their masses are 
considerably larger than those of ordinary hadrons, and that hyperquarks are 
much heavier than quarks. Hyperhadrons might have been 
produced in the Early Universe. However, 
because not even the lightest of these is observed today, 
the lightest hyperquark itself cannot be stable.
Consequently, a new class of massive bosons which 
generate hyperquark decays must exist. 
  
In this paper, we discuss the spectrum of composite spin 1 bosons 
that can be constructed in the preon model as well as their role 
in various weak decay processes. 
We assume that all massive composite bosons including the electroweak gauge 
bosons $W$ and $Z$ of the standard model, as well as those responsible 
for hyperquark decay, remain tightly bound at least up to $\cong 10^{16}$ GeV 
(grand unification scale). 
Explicit preon degrees of freedom do not appear below this scale 
so that the corresponding interactions between preon bound states 
can be described by approximate {\it effective} gauge theories. In particular, 
for low energies  $< 10^{3}$ GeV (Fermi scale) one recovers 
the standard model Lagrangian with left-right asymmetric weak interactions, 
while for $\cong 10^{9}$ GeV (partial unification scale), it is suggested that
an effective SU(6) gauge theory unifies left-right symmetric extended 
weak interactions including the new gauge bosons generating hyperquark 
decay with the hypercolor interaction. 

We also address the issue of the mass scale, where 
hyperquarks appear. The momentum dependence 
of the different gauge couplings is used to predict the energy 
scale where they converge and a unified gauge theory with
a single coupling  occurs. 
This in turn enables us 
to put limits on the nonperturbative regime of the hypercolor force 
and the mass of the lightest hypermeson. In short, 
the purpose of this work is to address the following questions:
\begin{itemize}
\item[(i)] How are massive weak gauge bosons 
described in the preon model? 
\item[(ii)] Which processes and gauge bosons are responsible for 
hyperquark instability? 
\item[(iii)] What is the effect of hyperquarks in various weak interaction 
processes? 
\item[(iv)] At what energy scale do hyperquarks and their composites
appear?
\end{itemize} 

Important questions such as gauge boson and
fermion mass generation are not discussed here. It may suffice to say that 
in the present model there are no fundamental Higgs fields although
there could be composite scalars such as hyperquark bound states. 
This is reminiscent of technicolor models and appears to be promising. 
However, closer inspection shows that hyperquarks are singlets under 
weak isospin. Therefore, they and their scalar bound states do 
not have the same SU(2)$_{W_{L}}\times $U(1)$_Y$ group structure as the 
standard model fermions and Higgs fields. Thus, hyperquark bound state scalars 
cannot give mass to any of the standard model gauge bosons and fermions. 
In this paper, we concentrate on the spectrum of vector bosons.
We hope to come back to the issue of scalar particles and mass generation
in the preon model elsewhere. 

The paper is organized as follows. Section~\ref{sec:fermionicboundstates} 
gives a short review 
of fermionic bound states in the preon model. 
In sect.~\ref{sec:extendedweak}, 
taking as straightforward a position as possible,
we discuss an extension of the standard electroweak theory,
making explicit the preon content of the massive electroweak 
gauge bosons including those responsible for hyperquark-quark transitions.  
A further generalization of the extended weak and the hypercolor interactions 
is discussed in sect.~\ref{sec:partialunificationscheme}. 
The resulting theory, which we call partial 
unification, represents a necessary step towards grand unification, as shown 
in sect.~\ref{sec:grandunification}.
Sect.~\ref{sec:runningcoupling} provides numerical results for the 
running couplings and unification constraints from where  mass ranges 
of the heavy bosons, the lightest neutrino, and the hyperquarks
are obtained. The predicted hyperhadron masses are within reach 
of the Large Hadron Collider at CERN.
Sect.~\ref{sec:summary} contains a summary and outlook. 

\section{Fermionic bound states}
\label{sec:fermionicboundstates}

In the Harari-Shupe model~\cite{har79,har80a,har81a} all quarks and leptons 
are built from just two spin-$\frac{1}{2}$ fermions (preons).
According to Harari and Seiberg~\cite{har80a} the two types of preons
belong to the following representations of the underlying exact gauge group
SU(3)$_{H}$ $\times$ SU(3)$_{C}$ $\times$ U(1)$_{Q}$: T: (3,3)$_{\frac{1}{3}}$ 
and V: (3,$\bar{3}$)$_{0}$,
where the first (second) argument is the dimension of the representation in
hypercolor (color) space and the subscript denotes the electric charge $Q$.
The fundamental Lagrangian of the preon model~\cite{har81a} reads
\bea
\label{preonlagrangian}
{\cal L} & = & 
\bar{T} \, \left ( \slashed{\partial} + g_{H} \, \slashed{A}_{H} + g_{C} \, 
\slashed{A}_C 
+\frac{1}{3}\, e \, \slashed{A}_{Q} \right )T \, \nonumber \\ 
& + &
\bar{V} \, \left ( \slashed{\partial} + g_{H} \, \slashed{A}_{H} + g_{C} \, 
\slashed{A}_{C} 
\right ) \,  V \nonumber \\
& & -\frac{1}{4} F_{H}\, F_{H} -\frac{1}{4} F_{C}\, F_{C} -
\frac{1}{4} F_{Q}\, F_{Q}, 
\eea
with $\slashed{A}=\gamma_{\mu} A_{a}^{\mu} \lambda^{a}$ representing the three
fundamental gauge fields of the theory with their respective
coupling strengths $g_{H}$, $g_{C}$, and $e$. 
The last three terms in Eq.(\ref{preonlagrangian}) 
represent the kinetic energies of the gauge fields, where the 
field strength tensors $F$ are as usual given in terms of the $A^{\mu}$.  
The $\gamma_{\mu}$ are the Dirac matrices and $\lambda^{a}$ are generators 
of the corresponding gauge groups.

Although there are two degenerate types of preons (T and V) there is 
no global SU(2) isospin symmetry on the preon level because the charged and 
neutral preon belong to different representations in color space.  
To be consistent with the 
parity assignment for the standard model
fermions, the intrinsic parity $\Pi$ of the $T$ and $V$ preons 
must be different~\cite{abms05}. The different parities of $T$ and $V$ 
preons provide a possible explanation for parity violation 
at low energies as will be 
discussed in sect.~\ref{sec:partialunificationscheme}.

\begin{table}[htb]
\begin{tabular}{l c c c c c c}
\hline
preon & H & C & Q & ${\cal P}$ & $ \Upsilon $ & $\Pi$ \\
\hline
$T$ & $3$ & $3$ & $+\frac{1}{3}$ & $+\frac{1}{3}$ & $+\frac{1}{3}$ & $-1$ \\
\hline
$V$ & $3$ & $\bar{3}$ & $0$ & $+\frac{1}{3}$ & $-\frac{1}{3}$ & $+1$  \\    
\hline
$\bar{V}$ & $\bar{3}$ & $3$ & $0$ & $-\frac{1}{3}$ & $+\frac{1}{3}$ & $-1$ \\
\hline
$\bar{T}$ & $\bar{3}$ & $\bar{3}$ & $-\frac{1}{3}$  & $-\frac{1}{3}$ & 
$-\frac{1}{3}$ & $+1$ \\
\hline
\end{tabular}
\caption{The color $C$, hypercolor $H$, electric charge $Q$, preon number
${\cal P}$, $\Upsilon$ number, 
and intrinsic parity $\Pi$  of preons and antipreons 
(see also~\cite{abms05}).} 
\label{tab:preon-entities}
\end{table}

Preons and their bound states are characterized by new quantum numbers
~\cite{har80}. These are the preon number ${\cal P}$ and $\Upsilon$ number,
which are linear combinations of the numbers of T-preons n(T) and 
V-preons n(V) in a given state
\begin{eqnarray}
\label{preonnumber}
{\cal P} &=& \frac{1}{3} \left(n(T) + n(V)\right) \\
\Upsilon &=& \frac{1}{3} \left(n(T) - n(V)\right). \nonumber
\end{eqnarray}
The factor $\frac{1}{3}$ in Eq.(\ref{preonnumber}) is a convention.
The $\Upsilon$ number is also related to the baryon (B) and lepton (L) 
numbers of the standard
model as $\Upsilon = B - L$.
The antipreon numbers $n(\bar{T}$) and $n(\bar{V}$) are defined as
$n(\bar{T}) = - n(T)$ and $n(\bar{V}) = - n(V)$.
The preon quantum numbers of individual preons are summarized 
in Table~\ref{tab:preon-entities}.

There is a connection between the ${\cal P}$ and $\Upsilon$ numbers, and
the electric charge $Q$ of the preons
\begin{equation}
\label{gellmannnishijima}
Q = \frac{1}{2} \left({\cal P} + \Upsilon\right),
\end{equation}
which can be readily verified from Table~\ref{tab:preon-entities}.
This generalized Gell-Mann-Nishijima relation does not only hold for the 
preons but for all bound states, such as  
leptons, quarks, hyperquarks and their bound states, as 
well as the effective weak gauge bosons.
\begin{table*}[htb]
\begin{center}
\begin{tabular}{ l c c c c c c c c c }
\hline
state & preon content & bound state &  ${\cal{P}}$ & $\Upsilon$ &  B & L &
$Q$ & $\Pi$ & $T_3$ \\ 
\hline
 &$\left( VVV \right)$ & $\left( \nu_{e}, \nu_{\mu}, \nu_{\tau} \right)$ & 
$+1$ & $-1$ & $0$ & $+1$ & $0$ & $+1$ & $+\frac{1}{2}$  \\
leptons & & & & & & & & & \\
 &$\left(\bar{T}\bar{T}\bar{T} \right)$ & $\left( e^{-}, \mu^{-}, \tau^{-} 
\right)$ & 
$-1$ & $-1$ & $0$ & $+1$ & $-1$ & $+1$ & $-\frac{1}{2}$  \\
\hline
 &$\left( TTV \right)$ & $\left( u, c, t \right)$  & $+1$ & $+\frac{1}{3}$ & 
$+\frac{1}{3}$ & 0 & $+\frac{2}{3}$ & $+1$ & $+\frac{1}{2}$     \\
quarks & & & & & & & & &  \\
 &$\left(\bar{T}\bar{V}\bar{V} \right)$ & $\left( d, s, b \right)$ &  $-1$ & 
$+\frac{1}{3}$ & $+\frac{1}{3}$ & $0$ & $-\frac{1}{3}$ & $+1$ & 
$-\frac{1}{2}$ \\ 
\hline
 &$\left( TT\bar{V} \right)$ & $\left( \tilde{u},\tilde{c},\tilde{t} 
\right)$ & $+\frac{1}{3}$ & $+1$ & $+1$ & $0$  & $+\frac{2}{3}$ & $-1$ & 0 \\  
hyperquarks & & & & & & & & & \\
 &$\left( \bar{T}VV \right)$ & $\left( \tilde{d}, \tilde{s}, 
\tilde{b} \right)$ & 
$+\frac{1}{3}$ & $-1$ & $-1$ & $0$  & $-\frac{1}{3}$ & $+1$ & 0 \\
\hline
\end{tabular}  
\caption{Allowed three-preon bound states representing leptons, quarks and
hyperquarks and their quantum numbers (see also~\cite{abms05}). Formally, the
hyperquarks are obtained from the corresponding quarks by interchanging: 
$ V \leftrightarrow \bar{V}$ (hyperquark transformation).}
\label{tab:all fermions}
\end{center}
\end{table*} 

As shown in Ref.~\cite{abms05} the 't Hooft anomaly condition, 
which demands that the anomalies on the preon level match those 
on the bound state level, can be satisfied if in addition to leptons 
and quarks a third  fermionic bound state type, called hyperquarks, 
is introduced.  Hyperquarks have the same electric charge as the 
corresponding quarks.  However, instead of being color triplets 
and hypercolor singlets as ordinary quarks, 
they are color singlets and hypercolor triplets. 
Moreover, because of the different parities of $\tilde{u}$
and $\tilde{d}$ one cannot define a weak 
SU(2) isospin symmetry for hyperquarks. Therefore, they 
do not participate in the usual left-right asymmetric 
weak interaction but must couple left-right symmetrically calling 
for a left-right symmetric extension of weak interactions.

The same conclusion is also obtained from the anomaly freedom
constraint of this extended electroweak theory. 
In the standard model the anomaly contributions of quarks and leptons 
cancel. There are no additional fermionic bound states that could cancel 
an anomaly contribution coming from hyperquarks. Therefore, hyperquarks 
must not contribute to electroweak anomalies. The only way this 
can be achieved is that their weak interactions be left-right symmetric.
Formally, hyperquarks are obtained from quarks by replacing their 
neutral preons with their antiparticles. We refer to this process 
as hyperquark transformation.

In summary, we can construct two groups of preon bound states 
(leptons and quarks) with the same intrinsic parities and
integer preon number for which a new quantum number 
``weak isospin'' corresponding to an effective 
SU(2) chiral isospin symmetry can be defined in terms of the 
preon number ${\cal P}$ as
\be
T_{3} =  \frac{1}{2}\, {\cal P}. 
\ee
The members of the third group (hyperquarks) necessarily have opposite 
intrinsic parities and fractional preon number and thus do not possess 
weak isospin.
The three different types of fermionic bound states and their quantum numbers
are shown in Table~\ref{tab:all fermions}.

\section{Bosonic bound states and extended weak interactions}
\label{sec:extendedweak}
Hyperquarks are hypercolored objects and thus cannot exist as free particles 
but must be confined into hypercolorless 
bound states such as hypermesons and hyperbaryons. This is 
in complete analogy to quarks being confined into colorless mesons and 
baryons. Although hyperhadrons might have been created at 
sufficiently high energies available in the Early Universe they are no longer 
observed today and hence cannot be stable. This leads to the question 
which gauge bosons are responsible for their decay.

We begin our discussion with charged
hypermeson decay, which is generated by a new gauge boson $\tilde{W}$ followed
by a short exposition of the charged and neutral weak transitions
between fermionic bound states. In section~\ref{subsec:hyperquarkdecay} 
we discuss the issue of hyperquark decay and show that it is 
mediated by a six-preon bosonic bound state, called $\chi$,
which can be thought of as two-neutrino bound state.  
The scenario of lepton number violating neutrino-antineutrino
oscillations characteristic of Majorana neutrinos 
is described in section~\ref{subsec:oscillation}. There, 
also the consequences for the formulation of an extended left-right symmetric 
weak interaction theory are expounded.

\subsection{Weak meson and hypermeson decays into leptons}
\label{subsec:hypermeson}
\begin{widetext}
\begin{figure*}
\resizebox{1.0\textwidth}{!}{
\includegraphics{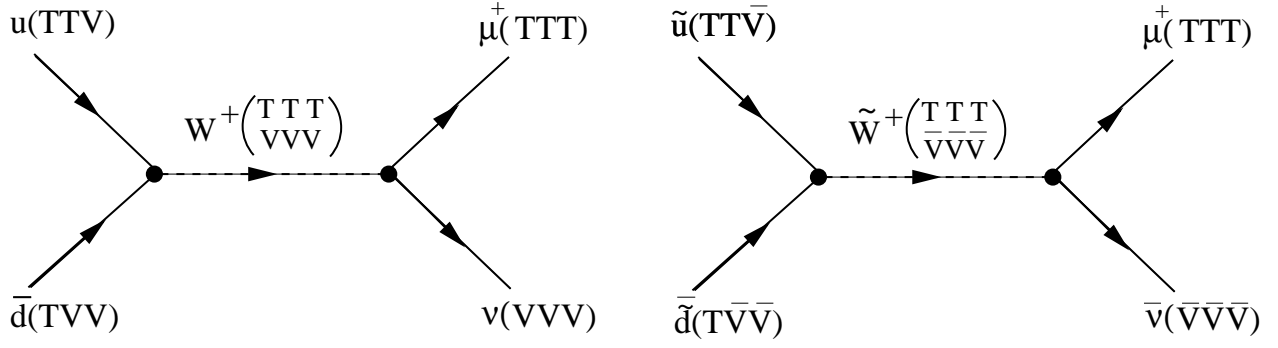}}
\caption{\label{figure:meson_decay} Weak pion decay into leptons (left) 
and lepton number violating weak hyperpion decay (right) 
mediated by bosonic preon bound states.}
\end{figure*}
\end{widetext}

As stated in the introduction and shown in Fig.~\ref{figure:meson_decay}, 
in the preon model the weak decays of hadrons and 
hyperhadrons are mediated by composite gauge bosons.
For example, the weak decay of the 
positively charged $\pi$-meson into a muon and a neutrino can be 
schematically written as


\begin{eqnarray}
\label{mesondecay}
&\pi^{+}& \left(\begin{array}{c} u \left(T T V \right) 
\\ \bar{d} \left(T V V \right) \end{array}\right)
\longrightarrow
W^{+}\left(\begin{array}{c} T T T \\ 
V V V \end{array}\right)  \nonumber \\
& \longrightarrow &  \mu^{+}\left(T T T \right) 
+\nu \left(V V V\right). 
\end{eqnarray}
  
As is obvious from the notation, 
the compositeness of quarks and leptons implies that the weak gauge 
bosons are composed of six preons~\cite{har81a}, and furthermore 
that initial and final states correspond merely to different arrangements 
of these preons. 

Similarly, the corresponding hyper-$\pi$-meson decay reads 
\begin{eqnarray}
\label{hypermesondecay}
&\tilde{\pi}^{+}& \left(\begin{array}{c} \tilde{u}
\left(T T \bar{V}\right) 
\\ {\bar{\tilde{d}}}\left(T\bar{V} \bar{V}\right) \end{array}\right)
\longrightarrow
\tilde{W}^{+}\left(\begin{array}{c} T T T \\ \bar{V}\bar{V}\bar{V}  
 \end{array}\right)  \nonumber \\
&\longrightarrow&  \mu^{+}\left(T T T \right) 
+\bar{\nu} \left( \bar{V}\bar{V}\bar{V} \right). 
\end{eqnarray}
A comparison of the preon content of the bound states in
Eq.(\ref{hypermesondecay}) and Eq.(\ref{mesondecay}) shows
that the neutral preons $V$ are replaced by their antiparticles $\bar{V}$ 
when going from quarks to hyperquarks~\cite{abms05} or from mesons to 
hypermesons (hyperquark transformation). 
In addition, in both cases the intermediate gauge bosons are formed by a mere
rearrangement of the initial state preons. Hence, we propose that  
the existence of left-right symmetrically coupling hyperquarks 
entails the existence of 
a new class of composite weak gauge bosons, called $\tilde{W}$, which 
couple left-right symmetrically to fermions.

Although there is a certain analogy between weak meson and 
hypermeson decays into leptons there is an important difference
between them. According to the quantum number assignments in 
table~\ref{tab:all fermions},
the latter process simultaneously violates lepton and baryon number
($ \Delta B = \Delta L = -2$), which indicates that it occurs only 
at a higher energy scale where left-right symmetry is restored.
In sect.~\ref{sec:runningcoupling} this scale is calculated as 
$M_{P}\cong 10^{9}$ GeV. Note that in both processes 
the preon ${\cal P}$ and $\Upsilon=B-L$ numbers remain conserved.

\subsection{Charged and neutral weak transitions 
between fermionic preon bound states}
\label{subsec:chargedandneutral}
In the preon model, the weak transitions among the members of quark 
and lepton weak isospin doublets caused by the charged weak currents 
of the standard model are written as
\begin{eqnarray}
\label{quarkleptondecay}
 u\left(TTV\right)\!\! & \longrightarrow & 
\!\!d\left(\bar{T}\bar{V}\bar{V}\right)
+ W^{+}\left(\begin{array}{c} TTT \\ VVV \end{array}\right) \nonumber \\
e^{-}\left(\bar{T}\bar{T}\bar{T}\right)\!\! 
& \longrightarrow & \!\!\nu\left(VVV\right)
+ \!W^{-}\left(\begin{array}{c} \bar{T}\bar{T}\bar{T} \\ \bar{V}\bar{V}\bar{V} 
\end{array}\right)\!\!. 
\end{eqnarray}
In these transitions all three 
preons of the final fermionic bound states are created  
from the vacuum, while the initial state preons and the antiparticle 
counterparts of the three vacuum pairs merge to form the weak gauge bosons.  

Analogously, the corresponding weak transitions among hyperquarks 
imply the existence of new gauge bosons, $\tilde{W}$,
which may then also generate transitions within lepton doublets 
\begin{eqnarray}
\label{tildedecay}
\tilde{u}\left(TT\bar{V}\right)\!\! & \longrightarrow &
\!\!\tilde{d}\left(\bar{T}VV\right)
+ \tilde{W}^{+} \left(\begin{array}{c} TTT \\ 
\bar{V}\bar{V}\bar{V} \end{array}\right)  \nonumber \\
e^{-}\left(\bar{T}\bar{T}\bar{T}\right)\!\! 
& \longrightarrow & \!\!\bar{\nu}\left(\bar{V}\bar{V}\bar{V}\right)
+ \tilde{W}^{-}\left(\begin{array}{c} \bar{T}\bar{T}\bar{T} 
\\ VVV \end{array}\right)\!\!.
\end{eqnarray}
In contrast to the usual electron-neutrino transition in 
Eq.(\ref{quarkleptondecay}),
the $\tilde{W}$ induced process in Eq.(\ref{tildedecay}) leads to 
an antineutrino in the final state and thus violates lepton number 
conservation ($\Delta L=-2$). On the other hand, $\Upsilon=B-L$ 
and ${\cal P}$ are conserved because $\Upsilon=-2$ 
and ${\cal P}=0$ for the $\tilde{W}^{-}$ boson according 
to Table~\ref{tab:low6preon}.
Similarly, in the hyperquark sector the charged weak transition
violates baryon number ($\Delta B=-2$) but again $\Upsilon=B-L$ 
and ${\cal P}$ are conserved.
One also notices that these charged weak transitions 
leave the type of fermionic bound state invariant, i.e., 
a quark remains a quark, a hyperquark remains a hyperquark,
and a lepton remains a lepton. 

Because neutral currents do not change the internal quantum numbers 
of the fermionic bound states involved in the transition, 
the $W_0$ boson must be a linear combination of the two neutral six-preon 
states. For the neutral standard model gauge bosons we define
\bea
\label{Zboson}
W_{0} & = & \frac{1}{\sqrt{2}} \left[
\left(\begin{array}{c} \bar{T}\bar{T}\bar{T} \\ TTT \end{array}\right)- 
\left(\begin{array}{c} \bar{V}\bar{V}\bar{V} \\ VVV \end{array}\right) 
\right] \nonumber \\
B_{0} & = & \frac{1}{\sqrt{2}} \left[
\left(\begin{array}{c} \bar{T}\bar{T}\bar{T} \\ TTT \end{array}\right)+ 
\left(\begin{array}{c} \bar{V}\bar{V}\bar{V} \\ VVV \end{array}\right) 
\right]. 
\eea
Note that applying the hyperquark transformation to Eq.(\ref{Zboson}) leaves
these states invariant so that we need not introduce 
additional neutral bosons $\tilde{W_{0}}$ and $\tilde{B_{0}}$.
As a linear combination of the two states in Eq.(\ref{Zboson}) the
$Z$-boson is a pure $(3V,3{\bar V})$ state, while the orthogonal
combination $(3T,3{\bar T})$ state can in some sense 
be interpreted as an {\it effective} photon similar to the vector
meson dominance model. In this way the standard model weak forces are seen 
to be mediated by {\it effective} gauge bosons composed of six preons.

As far as we can see, 6 $T$ and 6 $\bar{T}$ 
states do not occur in any weak process and are not considered here.
The same applies to six-preon bound states consisting 
of one charged preon and five neutral ones or one neutral preon and five
charged ones. 
In Table~\ref{tab:low6preon} we list the composite gauge bosons introduced
so far as well as their preon content and quantum numbers.

In summary, for the weak gauge bosons $W$, $Z$, and $\tilde{W}$, 
the following rules apply: 
(i) leptons and quarks transform via $W$ exchange; (ii) leptons and 
hyperquarks transform via $\tilde{W}$  exchange; 
(iii) leptons, quarks, and hyperquarks 
transform via $Z$ exchange. But neither the standard model gauge bosons
nor the newly introduced $\tilde{W}$ can transform hyperquarks into quarks.
\begin{table}[h]
\begin{center}
\begin{tabular}{l c c c c c c}
\hline
state & content &  ${\cal{P}}$ & $\Upsilon$ &  $Q$ & $\Pi$ & $T_3$ \\ 
\hline
$ W^{-}$ & $(3\bar{T}, 3\bar{V})$ & 
$ -2 $ & $ 0 $ & $-1$ & $-1$ & $-1$ \\ 
\hline
$W^{+} $ & $(3T, 3V)$ & $ +2 $ & $ 0 $ &$ +1$ & $-1$ & $+1$ \\
\hline  
$W_{0}/B_{0} $  & $(3(T\bar{T}), 3(V\bar{V}))$ & $ 0 $ & $ 0 $ & $ 0 $ &$-1$
& $0$\\ 
\hline\hline
$\tilde{W}^{-} $ & $(3{\bar T},3V)$   
& $ 0 $ & $ -2 $ & $ -1$ & $+1$ & $0$ \\
\hline
$\tilde{W}^{+} $ & $(3 T,3\bar{V})$  
& $ 0 $ & $ +2 $ & $ +1$ & $+1$ & $0$\\
\hline
\end{tabular}  
\caption{Preon content and quantum numbers of the 
standard model weak gauge bosons 
$W$ and $B_{0}$ and the newly introduced gauge bosons $\tilde{W}$.}
\label{tab:low6preon}
\end{center}
\end{table} 

\subsection{Weak hyperquark decays into quarks}
\label{subsec:hyperquarkdecay}
\begin{figure}
\resizebox{0.4\textwidth}{!}{
\includegraphics{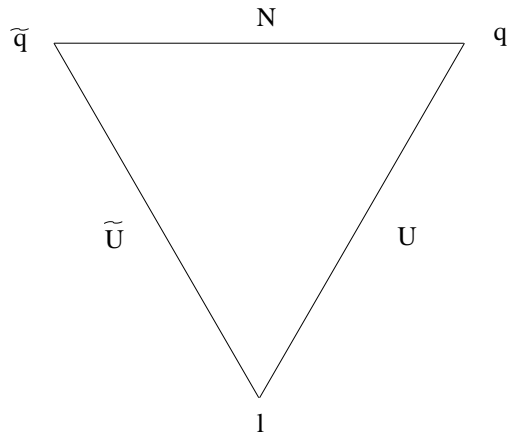}}
\caption{\label{figure:fermion_triangle} Fermion triangle.
The three fermionic bound states, hyperquarks $(\tilde{q})$, 
quarks $(q)$, and leptons $(l)$  
and their antiparticles are placed at the corners of a triangle. 
The dipreonic bound states $(N, U, \tilde{U})$ 
and their antiparticles describing transitions between these fermions 
are placed along the edges.}
\end{figure}

Similar to the case of hypermesons discussed in sect.~\ref{subsec:hypermeson},
the nonobservation of hyperbaryons 
implies that they are not stable. Consequently, there must 
be transitions from hyperquarks to known fermionic preon bound states.
In principle, hyperquarks can decay into quarks and into leptons.  
However, below the grand unification (GUT) scale the former process dominates
for the following reason. The decay of quarks into leptons, 
as required for proton decay is suppressed due to the heavy mass of 
the GUT gauge bosons of the order of $10^{16}$ GeV, 
which corresponds to lifetimes of the order of $10^{35}$ y. 
Because the preonic substructure of 
hyperquarks and quarks are very similar,
transitions from hyperquarks 
to leptons are also suppressed at lower energies.
The decay of quarks and hyperquarks into leptons mediated by dipreonic 
bound states $U$ and ${\tilde U}$ occuring at the GUT scale 
will be considered in section~\ref{sec:grandunification}. 
Clearly, the fact that hyperquarks are not observed today
requires that their lifetime be much shorter than the lifetime of the Universe
($10^{10}$ y), and that the gauge bosons $N$ responsible for 
hyperquark decay into quarks be lighter than the GUT bosons.
The transitions between the three types of fermionic bound states
mediated by dipreonic bound states $N$, $U$, and ${\tilde U}$ 
are shown in Fig.~\ref{figure:fermion_triangle}.

In our previous paper~\cite{abms05}, we have seen that the 
transition from hyperquarks to quarks is formally obtained by interchanging 
the neutral preon by its antiparticle
($V \leftrightarrow \bar{V} $). 
Hyperquark decays into quarks are generated via a 
new class of electrically neutral bosons, 
called $N$-bosons or neutralons $N\left(VV\right)$ 
and $\bar{N}\left(\bar{V}\bar{V}\right)$
\begin{eqnarray}
\label{hyperquarkneutralon}
\tilde{u}\left(TT\bar{V}\right) & \longrightarrow & u\left(TTV\right) + 
\bar{N}\left(\bar{V}\bar{V}\right) \nonumber \\
\tilde{d}\left(\bar{T}VV\right) & \longrightarrow & 
d\left(\bar{T}\bar{V}\bar{V}\right) + 2 N\left(VV\right).
\end{eqnarray}

As color and hypercolor triplets, neutralons are confined particles. 
In order to carry away the energy and momentum made available 
in the $\tilde{q} \to q$ transition, an unconfined neutral particle
must be emitted. Such a hypercolor and color singlet  
can be formed by an $N \bar{N}$ pair or three neutralons.
The emission of two $N \bar{N}$ pairs, 
as occuring in hyperproton decay according to
Eq.(\ref{hyperquarkneutralon})  
seems at first sight possible. But an $N \bar{N}$ state has 
neither weak isospin
nor electric charge and thus it cannot couple to any of the known low energy 
bosons such as $Z$ or $\gamma$ into which it could annihilate. 
Therefore, direct hyperproton decay via the emission of two 
$N \bar{N}$ pairs does not work. However, annihilation into the vacuum  
is possible for {\it three} $N \bar{N}$ pairs. 
In fact, $3N$ or $3 \bar{N}$(and thus 3 $N \bar{N})$ form colorless and 
hypercolorless bound states which can decay into neutrinos 
and antineutrinos. In the following the $3N$ and $3\bar{N}$ states 
are called $\chi$ and $\bar{\chi}$.

More generally, we postulate that in all weak processes 
occuring below the grand unification scale of $10^{16}$ GeV, 
such as hyperquark-quark transitions, as well as in 
processes involving $W$, $\tilde{W}$ 
and $Z$ exchange, 
the generation or annihilation of preon-antipreon pairs is only possible 
for integer multiples of three preon-antipreon pairs. 
This is symbolically written as
\begin{equation} 
\label{modulo3rule}       
n_{(\bar{T}T)} + n_{(\bar{V}V)} = 3 k 
\quad \left(\mbox{for E} \leq 10^{16}\, \mbox{GeV}\right), 
\end{equation}
where $n_{(\bar{T}T)}$, $n_{(\bar{V}V)} $, and $k$ are natural numbers.
Thus, in any weak interaction below the grand unification scale, 
only six-preon (see previous subsection) or three-dipreon bosons 
(see next subsection) are involved. 
We refer to this as ``preon triality rule''.

The simplest hyperbaryon decay process seems to proceed 
via a $\tilde{{\Delta}}$ type hyperbaryon in which the six-preon bound states 
$\chi$ and ${\bar \chi}$ are produced
\begin{eqnarray}
\label{dn}
\tilde{\Delta}^{++}
\left(\tilde{u}\tilde{u}\tilde{u}\right) & 
\longrightarrow & \Delta^{++} \left(uuu\right) + \bar{\chi} 
\nonumber \\
\tilde{\Delta}^{-}
\left(\tilde{d}\tilde{d}\tilde{d}\right) & \longrightarrow & 
\Delta^{-}\left(ddd\right) + 2 \chi, 
\end{eqnarray}
and where the latter subsequently 
decay into two neutrinos ($\chi \to 2 \nu$) 
or antineutrinos ($\bar{\chi} \to 2 \, \bar{\nu}$). 
Thus, the transition from hyperquarks to quarks 
is only possible within bound states of three hyperquarks of the same 
charge state ($\tilde{u},\tilde{c},\tilde{t}$ or 
$\tilde{d},\tilde{s},\tilde{b}$).
Other baryonic hyperquark bound states,  
$\tilde{\Delta}^{+} \left(\tilde{u}\tilde{u}\tilde{d}\right)$ and 
$\tilde{\Delta}^{0} \left(\tilde{u}\tilde{d}\tilde{d}\right)$, 
decay via a two-step process,
i.e., they first change into a 
$\tilde{\Delta}^{++}(\tilde{u}\tilde{u}\tilde{u})$
and $\tilde{\Delta}^{-}(\tilde{d}\tilde{d}\tilde{d})$ 
via $\tilde{W^{\pm}}$ emission as in 
Eq.(\ref{tildedecay}) and then decay into ordinary baryons 
via $\chi$ emission as in Eq.(\ref{dn}).  Because of the vanishing isospin
of hyperhadrons the spin-symmetric $\tilde{\Delta}$ baryons are 
the lightest fermionic hyperhadron states.

\subsection{Left-right symmetric weak interactions and
neutrino-antineutrino oscillations}
\label{subsec:oscillation}

In this section we discuss in more detail how the 
existence of hyperquarks leads to a left-right symmetric extension
of standard model weak interactions. 
As explained before, hyperquarks do not have weak isospin
and therefore they couple left-right symmetrically to the new gauge 
bosons $\tilde{W}$ and $N$. At the partial unification scale $M_P$, 
where these new left-right symmetric gauge bosons appear, 
hyperquarks decay into quarks as in Eq.(\ref{hyperquarkneutralon}).
Therefore, at this scale quarks must also couple left-right symmetrically 
to the standard model weak gauge bosons $W$. This entails an extension 
of the standard model weak isospin group SU(2)$_{W_{L}}$  
to the gauge group SU(2)$_{W_{L}}\times$ SU(2)$_{W_{R}}$.

At the same energy, the effective gauge boson masses 
$M_{W_R}$, $M_{W_L}$, $M_{\tilde W}$  are of order $M_{P}$,
and the left-right symmetric gauge bosons can transform into each other. 
In particular, the $W$-bosons can change into the $\tilde{W}$-bosons 
and vice versa, 
\begin{eqnarray}
\label{wintowtilde}
\label{wtilde}
\tilde{W}^{+} \left(\begin{array}{c} TTT \\ \bar{V}\bar{V}\bar{V} 
\end{array}\right)
& \longleftrightarrow & W^{+}\left(\begin{array}{c} TTT \\ VVV \end{array}
\right)_{L/R} \nonumber\\  
\tilde{W}^{-} \left(\begin{array}{c} \bar{T}\bar{T}\bar{T} \\ VVV 
\end{array}\right)
& \longleftrightarrow & W^{-}\left(\begin{array}{c} \bar{T}\bar{T}\bar{T} \\ 
\bar{V}\bar{V}\bar{V} \end{array}\right)_{L/R}.
\end{eqnarray}
In addition, there are transitions between the $\chi$, $Z$, 
and $\bar{\chi}$ bosons 
\begin{equation}
\label{chiz}
\chi\left(\nu \, \nu \right) \longleftrightarrow 
Z\left( \nu \, \bar{\nu} \right)  \longleftrightarrow 
\bar{\chi}\left ( \bar{\nu}\, \bar{\nu} \right).
\end{equation}
The transformations in Eq.(\ref{wintowtilde}) and Eq.(\ref{chiz}) are 
a reflection of left-right symmetry restoration at the partial 
unification scale.
\begin{figure}
\resizebox{0.49\textwidth}{!}{
\includegraphics{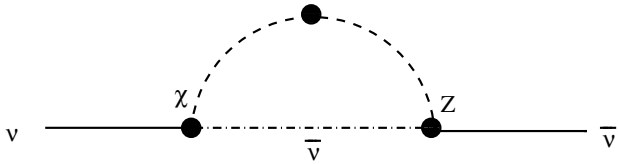}}
\caption{\label{figure:neutrino_antineutrino} Neutrino-antineutrino
oscillation via $\chi$ and $Z$ emission and absorption.}
\end{figure}
These left-right symmetric interactions also enable transitions between 
the left- and right-handed neutrino sectors 
\begin{equation}
\label{nlrsy}
\nu \longleftrightarrow \bar{\nu} \quad 
\ {\rm or} \  \quad \nu_{L} \longleftrightarrow \nu_{R}
\end{equation}
characteristic for two-component Majorana neutrinos~\cite{cas57,pat74},
for which the difference between neutrino and antineutrino 
is only one of helicity, i.e. $\nu_{L} \equiv \nu$ and
$\nu_{R} \equiv \bar{\nu}$. Such neutrino-antineutrino transitions 
can be thought of as proceeding via 
$\chi\leftrightarrow Z \leftrightarrow \bar{\chi}$ 
transitions as depicted in Fig.~\ref{figure:neutrino_antineutrino}.
Obviously, at this energy scale the inner parity and weak isospin 
of preon bound states is no longer a good quantum number.

In the preon model neutrino-antineutrino oscillations 
may be allowed for the following reasons. 
As discussed in sect.~\ref{sec:fermionicboundstates},  
the neutral $V$ and $\bar{V}$ preons have different 
$SU(3)_{C} \times SU(3)_{H}$ and parity assignments 
(see Table~\ref{tab:preon-entities}). Because none of the fundamental 
gauge interactions of the theory, namely SU(3)$_{C}$, SU(3)$_{H}$, 
and U(1)$_{Q}$ violates parity conservation, there can be no direct
transitions between the neutral $V$ preon and its antiparticle $\bar{V}$.
For the same reason, transitions between the neutralons 
$N(VV)$ and $\bar{N}(\bar{V}\bar{V})$ are forbidden.
However, for bound states of three $V$ and three
$\bar{V}$ preons, which are electrically neutral 
color and hypercolor singlets, there is from the viewpoint 
of the fundamental gauge symmetries no distinction between 
$\nu(VVV)$ and $\bar{\nu}(\bar{V}\bar{V}\bar{V})$ as pointed out by
Harari and Seiberg~\cite{har80}. 

As one can readily see from Table~\ref{tab:preon-entities},
the additive ${\cal P}$ and $\Upsilon$ quantum numbers are 
simultaneously violated in Eq.(\ref{nlrsy}) 
but in such a way that the total electric charge associated 
with the fundamental U(1)$_{Q}$ gauge interaction remains conserved
in accordance with the generalized 
Gell-Mann Nishijima relation Eq.(\ref{gellmannnishijima}), i.e.,
$\Delta Q = 0$ which entails $\Delta {\cal P} = - \Delta \Upsilon$.
Furthermore, it is observed that this violation can only occur 
for integer values of these quantum number and processes
involving three $V$ or three ${\bar V}$, 
whereas fractional values of ${\cal P}$ and $\Upsilon$ 
are strictly conserved because of their connection with fundamental gauge 
interactions (see Table~\ref{tab:preon-entities}). 
Analogous to the ${\cal P}$ 
and $\Upsilon$ number violation associated
with the direct neutrino-antineutrino oscillation,
the intrinsic parity violation in Eq.(\ref{nlrsy}) 
only occurs at the level of bound states of at least three preons.

Within each chiral sector the ${\cal P}$ and $\Upsilon$ 
quantum numbers are conserved as required by the anomaly 
equations~\cite{abms05}.
For this reason, any violation of these quantum numbers is accompanied by a 
simultaneous change of the chiral sector, which is only possible for massive 
particles. 
The corresponding transition rate is determined 
by the ratio of the neutrino mass and the partial unification scale, 
where the right-handed electroweak gauge bosons occur~\cite{pat74,moh81}. 
The heavier these bosons, the smaller the helicity changing transition rate 
and therefore the neutrino rest mass. This corresponds to the so-called
see-saw mechanism~\cite{moh81,har87}
\begin{equation}
\label{seesaw}        
\frac{m_{e}^{2}}{M_{P}} = m_{\nu_{e}}, 
\end{equation}   
where $m_{\nu_{e}}$ and $m_{e}$ are the neutrino and electron mass,
and $M_{P}\cong 10^{9}$ GeV is the scale of left-right symmetry restoration.

The extended left-rigt symmetric electroweak interactions, 
which go hand in hand with $\nu - \bar{\nu}$ and $\chi - Z - \bar{\chi}$ 
oscillations as depicted in Fig.~\ref{figure:neutrino_antineutrino}, 
lead to new $\Upsilon=B-L$ number violating processes, 
such as for example, (i) neutrinoless hyperbaryon decay
where $\Upsilon$- and $B$- numbers are violated and lepton number $L$ 
is conserved and (ii) neutrinoless double beta decay
where $\Upsilon$- and $L$- numbers are violated and $B$ is conserved
\begin{eqnarray}
\label{hqq}
\tilde{\Delta}^{++}
\left(\tilde{u}\tilde{u}\tilde{u}\right) & \!\!\longrightarrow 
\!\!& \Delta^{++} 
\left(uuu\right) + Z; \quad  \Delta \Upsilon = \Delta B = - 2 \nonumber \\
\tilde{\Delta}^{-}\left(\tilde{d}\tilde{d}\tilde{d}\right) & 
\!\!\longrightarrow \!\! & 
\Delta^{-} 
\left(ddd\right) + 2 Z; \quad \Delta \Upsilon = \Delta B = 4 \nonumber \\  
2 n(udd) & \!\!\longrightarrow \!\!& 2 p^{+}(uud) + 2 e^{-}; \ 
 \Delta \Upsilon \!=\! - \Delta L\! =\! -2. \nonumber \\
\end{eqnarray}
The discovery of any one of these $B$ or $L$ violating decays would 
lend some support to the ideas developed here. 

\section{Partial unification}
\label{sec:partialunificationscheme}
Having motivated an extension of the 
standard model weak interaction that accounts for
hyperquark decay, left-right symmetric 
weak gauge bosons, and direct neutrino-antineutrino oscillations,
we study in this section further aspects of this extension.
In particular, we propose that the generalized weak interactions
are part of a larger unification scheme for weak and strong
interactions between preon bound states that includes  
(i) 7 left-right symmetric standard model weak 
gauge bosons $W_{L}$, $W_{R}$, $B^0$, transforming according to
a simply extended rank 3 gauge group 
SU(2)$_{W_L}\times$SU(2)$_{W_R}\times$U(1)$_Y$, 
(ii) 20 new left-right symmetric gauge bosons $\tilde{W}$, $N$, $\bar{N}$, 
(iii) 8 hypergluons,
altogether 35 gauge bosons (see Table~\ref{tab:gaugegroups})
characteristic of an effective SU(6) gauge group. 
In addition, we emphasize that in the present model it is possible
to connect the left-right symmetry of weak interactions at high 
energies and its breaking at low energies to the existence
of hyperquarks and the new weak gauge bosons.
Before this, we discuss the consequences of the new effective 
gauge interactions and of hyperquarks 
for the momentum transfer dependence of the electroweak coupling strengths
$\alpha_{W}$, $\alpha_{Y}$ and $\alpha_{Q}$ and calculate the corresponding
Weinberg angle. 

\subsection{Extended electroweak couplings}
\label{subsec:extension}
In the standard model the left-handed leptons and quarks 
form weak isospin doublets transforming according to the
gauge group SU(2)$_{W_{L}}$ whereas the right-handed leptons and quarks 
are isospin singlets  transforming only according to the 
gauge group U(1)$_{Y}$.
The connection between electric charge $Q_{i}$, 
weak isospin $T_{3i}$, and weak hypercharge $Y_{Wi}$ of particle $i$ 
is given by the Gell-Mann Nishijima relation
\begin{equation}
\label{gmn}
Q_{i} = T_{3i} + Y_{Wi},
\end{equation}
where the particle index $i$ stands for a member of the lepton 
or quark doublets. The corresponding quantum numbers are given 
in Table~\ref{tab:isospin}.
\begin{table}[h]
\begin{tabular}{l c c c c  | c  c c }
\hline
Fermion & $T_{3}$ & $Y_{W} $ & $ B $ & $ L $
& $T_{3}^2$ & $Y_{W}^2$ & $Q^2$ \\
\hline
$\nu_{L}$ & $ + \frac{1}{2} $ & $ - \frac{1}{2}$ & $ 0 $ & $ + 1 $  & 
$\frac{1}{4}$ &  $\frac{1}{4}$ &  $0$ \\
\hline
$e^{-}_{L}$ & $ - \frac{1}{2} $ & $ - \frac{1}{2}$ & $ 0 $ & $ + 1 $ &
$\frac{1}{4}$ &  $\frac{1}{4}$ &  $1$ \\    
\hline
$e^{-}_{R}$ & $ 0 $ & $ - 1 $ & $ 0 $ & $ + 1 $ & 
$0$ &  $ 1$ &  $1$ \\
\hline\hline
$u_{L}$ & $ + \frac{1}{2} $ & $ + \frac{1}{6}$ & $ + \frac{1}{3} $ & $ 0 $ &
$\frac{1}{4}$ &  $\frac{1}{36}$ &  $\frac{4}{9}$ \\
\hline
$u_{R}$ & $ 0 $ & $ + \frac{2}{3}$ & $ +\frac{1}{3} $ & $ 0 $ & 
$0$ &  $\frac{4}{9}$ &  $\frac{4}{9}$ \\
\hline
$d_{L}$ & $ - \frac{1}{2} $ & $ + \frac{1}{6}$ & $ + \frac{1}{3} $ & $ 0 $ & 
$\frac{1}{4}$ &  $\frac{1}{36}$ &  $\frac{1}{9}$ \\
\hline
$d_{R}$ & $ 0 $ & $ - \frac{1}{3}$ & $ + \frac{1}{3} $ & $ 0 $ & 
$0$ &  $\frac{1}{9}$ &  $\frac{1}{9}$ \\
\hline\hline
$\tilde{u}_{L/R}$ & $ 0 $ & $ + \frac{2}{3}$ & $ + 1 $ & $ 0 $ &
$0$ &  $\frac{4}{9}$ &  $\frac{4}{9}$ \\
\hline
$\tilde{d}_{L/R}$ & $ 0 $ & $ - \frac{1}{3}$ & $ - 1 $ & $ 0 $ & 
$0$ &  $\frac{1}{9}$ &  $\frac{1}{9}$ \\
\hline
\end{tabular}
\caption{Weak isospin, weak hypercharge, baryon, and lepton quantum number 
assignments of the left- and right-handed leptons, quarks and 
left-right symmetric hyperquarks.
The squares of weak isospin, hypercharge, and electric charge 
are also given.} 
\label{tab:isospin}
\end{table}

It has been shown in section~\ref{subsec:chargedandneutral}
that hyperquarks do not participate 
in the usual charged current interactions mediated by $W$ exchange.
Therefore, with respect to the standard electroweak gauge group, 
the left- and right-handed hyperquarks 
are isospin singlets $(T_{3i}=0)$ analogous
to the right-handed leptons and quarks. 
With these properties of the composite fermions at hand, 
we can now proceed and discuss how the coupling constants
and their ratios are affected by the presence of the hyperquarks.

First, we note the standard definitions of the three electroweak 
couplings $\alpha_{Q}$, $\alpha_{W}$ and $\alpha'$ and the relation 
between them~\cite{fri75,moh92}
\bea
\label{WEI} 
\frac{\alpha_{Q}}{\alpha_{W}} & := &\sin^2 \Theta_{W} 
= \frac{\sum_{i} T_{3\,i}^2}{\sum_{i} Q_{i}^2} \nonumber \\
\frac{\alpha_{Q}}{\alpha'} & := &\cos^2 \Theta_{W} 
= \frac{\sum_{i} Y_{Wi}^2}{\sum_{i} Q_{i}^2} \nonumber \\
\frac{1}{\alpha_{Q}} & = & \frac{1}{\alpha_{W}} + \frac{1}{\alpha'},
\eea
where the third equation follows from dividing the sum of the first two 
by $\alpha_{Q}$.  This implies the relation 
\begin{equation}
\label{gmnsq}
\sum_i Q_{i}^{2} = \sum_{i} T_{3i}^{2} + \sum_{i} Y_{Wi}^{2},
\end{equation}
which can also be obtained directly from the square of the Gell-Mann Nishijima 
relation Eq.(\ref{gmn}) because the sum (over all fermions) of the mixed 
product terms vanishes.

In must be noted that an equality of $\alpha_{W}$ and $\alpha'$ cannot
be obtained because $\sum T_{3i}^{2} \ne  \sum_{i} Y_{Wi}^{2}$. Therefore,
the following coupling $\alpha_{Y}$ is introduced
\be
\label{redefinition}
\frac{1}{\alpha'} = 
\frac{1}{\alpha_{Y}} \, \frac{\sum_{i} Y_{Wi}^{2}}{\sum_{i} T_{3i}^{2}} 
\ee
which, when inserted in Eq.(\ref{WEI}) gives
\begin{equation}
\label{EW}
\frac{1}{\alpha_{Q}} = 
\frac{1}{\alpha_{W}} + \frac{\sum_{i} Y_{Wi}^{2}}{\sum_{i} T_{3i}^{2}} \,
\frac{1}{\alpha_{Y}}.
\end{equation}
In the standard electroweak theory (without hyperquarks) we have at 
the  weak scale $M_{Z}=91.2 \, \mbox{GeV}$ from Eq.(\ref{EW})
\begin{equation}
\label{EW1}
\frac{1}{\alpha_{Q}} = \frac{1}{\alpha_{W}} + \frac{5}{3} \,
\frac{1}{\alpha_{Y}}.
\end{equation}

Above this scale, at a certain energy which will
be determined in section~\ref{sec:runningcoupling} 
as $m_{hq} \cong 26$ TeV, the additional contributions 
of the hyperquarks affect the sums in 
Eq.(\ref{WEI}) and Eq.(\ref{EW}) as discussed in the following.
First, the sum $T_{3i}^{2}$ over (left-handed) quarks and 
leptons remains unmodified  
\begin{widetext}
\begin{equation}
\sum_{i} T_{3iL}^{2} = N_{G}\, \left [ 
\left( \frac{1}{2}\right)^{2}_{\nu}  +  \left(-\frac{1}{2}\right)^{2}_{e} 
+ N_{C} \left( \left(\frac{1}{2}\right)^{2}_{u}  + 
\left(-\frac{1}{2}\right)^{2}_{d} \right) 
\right ], 
\end{equation}
because hyperquarks do not have weak isospin.
Here, the factor $N_{C}$ stands for the number of quark colors,
and $N_{G}$ is the number of generations.   
Second, the sum of the weak hypercharge squares is modified by the 
presence of hyperquarks as
\begin{eqnarray}
\label{upsilonsum_{hq}}
\sum_{i} Y_{Wi}^2 & = &
 N_{G}\, \Biggl [  
\left(-\frac{1}{2}\right)^2_{\nu_{L}} + \left( -\frac{1}{2}\right)^2_{e_{L}} 
+ \left (-1 \right )^2_{e_{R}} \nonumber \\ 
& + & N_{C} \left( \left(\frac{1}{6}\right)^2_{u_{L}}  + 
\left(\frac{1}{6}\right)^2_{d_{L}} 
 + \left(\frac{2}{3}\right)^2_{u_{R}} + \left(-\frac{1}{3}\right)^2_{d_{R}} 
\right) \nonumber \\ 
& + & N_{H} \left( \left(\frac{2}{3}\right)^2_{\tilde{u}_{L}} + 
\left(-\frac{1}{3}\right)^2_{\tilde{d}_{L}}  
+ \left(\frac{2}{3}\right)^2_{\tilde{u}_{R}} + 
\left(-\frac{1}{3}\right)^2_{\tilde{d}_{R}} \right) 
  \Biggr ], 
\end{eqnarray}
where $N_{H}$ is the number of hypercolors. 
Third, the sum over the squared charges is 
\begin{equation}
\sum_{i} Q_{i}^2  = 
2 \, N_{G}\, \Biggl [  
 \left( -1 \right)^2_{e} 
 + N_{C} \left( \left(\frac{2}{3}\right)^2_{u}  + 
\left(-\frac{1}{3}\right)^2_{d} \right)  
 + N_{H} \left( \left(\frac{2}{3}\right)^2_{\tilde{u}} + 
\left(-\frac{1}{3}\right)^2_{\tilde{d}} \right) 
  \Biggr ], 
\end{equation}
\end{widetext}
where the factor of 2 is due to equal contributions from 
left-handed fermion doublets and right handed fermion singlets, 
and numerically $N_{C} = N_{H} = N_{G} = 3 $ 
according to Ref.~\cite{abms05}.

Because hyperquarks carry electric charge and weak hypercharge 
but no weak isospin, their inclusion in Eq.(\ref{WEI}) leads to
the following Weinberg angle  
\be
\label{WEI_hyperquark} 
 \sin^2 \Theta_{W} = \frac{\alpha_{Q}}{\alpha_{W}} =
\frac{\sum_{i} T_{3\,i}^2}{\sum_{i} Q_{i}^2} =
\frac{3}{13}.
\ee
Thus, we have $\sin^2\Theta_{W}=3/13 \cong 0.231$, which is close 
to the experimental Weinberg angle 0.23119(14) 
at $M_{Z}= 91.2$ GeV~\cite{par08}.
We take this as an indication for the physical relevance of hyperquarks.
In the original preon model of Harari and Seiberg~\cite{har80a} 
the weak mixing angle was predicted as $\sin^2\Theta_{W}= 1/4$.

Due to left-right symmetry restoration at the partial unification scale 
$M_{P} \cong 10^9$ GeV discussed in sect.~\ref{subsec:oscillation} 
we have left-right symmetric 
couplings of the standard weak gauge bosons and the new gauge bosons 
$\tilde{W}$ and $\chi$ to the fermions.
Consequently, above this scale, the sum over the squared 
isospin quantum numbers for leptons and quarks must run over both 
left- and right- handed fermions
\begin{equation}
\label{lriso}
\sum_{i} T_{3i}^2=\sum_{i} T_{3iL}^2 + \sum_{i} T_{3iR}^2, 
\end{equation}
and hence is twice as large as in Eq.(\ref{WEI_hyperquark}).
Furthermore, at the energy scale of $M_{P} \cong 10^9$ GeV, 
where left-right symmetry is restored, the electroweak coupling constants 
will necessarily have 
equal strength (see Fig.~\ref{figure:scattering})
\begin{equation}
\label{EQWY}
\alpha_{W/Y} := \alpha_{W} = \alpha_{Y}.
\end{equation}
Using these conditions in Eq.(\ref{WEI_hyperquark}) one obtains then for the
Weinberg angle at $M_{P}$
\begin{equation}
\label{ALPU}
\sin^2 \Theta_{W} = \frac{\alpha_{Q}}{\alpha_{W/Y}} = \frac{6}{13},  
\end{equation}
which is twice the experimental value of $\sin^2\Theta_{W} $ 
at $M_{Z}= 91.2$ GeV. 
We note that for a simply extended weak gauge group 
SU(2)$_{W_L}\times$SU(2)$_{W_R}\times$U(1) without hyperquarks one 
gets $\sin^2\Theta_{W}= 3/4$.

The breaking of left-right symmetry in standard electroweak theory 
at the $M_{Z}$ scale leads to a smaller value for the Weinberg angle 
which is only half of the value at $M_{P}$ given by Eq.(\ref{ALPU}).
This change of the Weinberg angle is accompanied
by a mass shift from the left-right symmetrically coupling 
bosons at $M_{P}$ down to their low energy counterparts at 
the Fermi scale $M_{Z}$.

At the energy $M_{P}$, where left-right symmetry is restored, 
the existence of an additional SU(2)$_{W_{R}}$ gauge group 
leads to an extended electroweak group 
SU(2)$_{W_{L}} \times $SU(2)$_{W_{R}}\times$ U(1)$_{Y}$~\cite{har80}
which itself is embedded in an even larger partial unification group 
SU(6)$_P$ with a single coupling constant 
\begin{equation}
\label{PUW}
\alpha_{P} := 2 \, \alpha_{W/Y}.
\end{equation}
Here, $\alpha_{P}$ is to be evaluated at 
the scale of left-right symmetry restoration and 
the factor 2 is due to the ensuing strength doubling according
to Eq.(\ref{lriso}) and depicted by the short dotted line 
in Fig.~\ref{figure:scattering}. Further aspects of the 
unification of weak with hypergluon interactions will be discussed next.

\subsection{Partial unification group SU(6)$_{P}$}
\label{subsec:partialunificationgroup}

We have seen that on the level of preon bound 
states new effective gauge interactions emerge, 
e.g. those mediated by the $\tilde{W}$ and the $N$ bosons, 
and that they occur at an energy where a left-right symmetric 
extension of standard model weak interactions is required. 
However, we have not yet incorporated the new gauge bosons 
in an appropriately enlarged effective gauge group.  
This can be achieved by combining the extended weak  
and strong hypercolor gauge interactions. 
Such a combination is also suggested by the fact that the coupling constants 
of SU(3)$_{H}$ and SU(3)$_{C}$ always run in parallel 
(see Fig.~\ref{figure:scattering}) 
and cannot reach equal strength because both groups have the same structure. 
Consequently, one of these color groups has to be embedded in 
a larger group if one wishes to unify the strong and weak interactions 
between preon bound states. 
 
We propose a partial unification by embedding the 
extended electroweak and hypercolor interactions within a broken 
gauge group SU(6)$_{P} \supset $ SU(3)$_{H}  
\times $SU(2)$_{W_{L}} \times $SU(2)$_{W_{R}} \times $U(1)$_{Y}$.
Note that SU(6) is the simplest rank 5 group having the same rank as
its subgroups. At the partial unfication scale $M_{P}$, the following 
equality of coupling constants is required to hold 
\begin{equation}
\label{PUH}
\alpha_{P} := \alpha_{H}= 2\, \alpha_{W/Y},
\end{equation} 
where the last equality follows from Eq.(\ref{PUW}).  
The SU(6) group also contains the hypercolored neutralons and the 
$\tilde{W}^{\pm}$ bosons.
The $6 \times 6$ matrix representing the effective SU(6) gauge bosons is 
schematically shown below
\be
\label{pu_matrix}
G_P=\left(\begin{array}{c} \hspace{-1.8cm} G_{H} \hspace{+2.2 cm} N \\ 
\bar{N} 
\quad \left(\begin{array}{cc} W^{0}_{R} \qquad W^{+}_{R}
\qquad \tilde{W}^{+} \\
W^{-}_{R} \qquad W^{0}_{L} \qquad W^{+}_{L} \\  
\tilde{W}^{-} \qquad W^{-}_{L} \qquad B^{0}
\end{array}\right) \end{array}\right). 
\ee
This scheme includes $3 \times 3$ matrices for the hypergluons $G_{H}$ 
and the neutralons $N$ and $\bar{N}$ transforming according to
SU(3)$_{H}$.  For the latter only the hypercolor degree 
of freedom is considered when counting their multiplicity. 
The partial unifification group is then a unitary group of dimension 6, 
comprising 8 hypergluons, 9 left-right symmetric weak gauge bosons,
and 18 left-right symmetric dipreonic neutralons, 
in total 35 generators as discussed above.
The corresponding gauge bosons with their respective energy scales 
are listed in Table~\ref{tab:gaugegroups}. 
Note that the diagonal
generators $G_{H}$, $W_{R}^0$ and $W_{L}^0$ have a $B_{0}$ 
admixture~\cite{moh92}.

The relevant SU(6) representations for the fermionic preon  
bound states arise from the direct product of three fundamental
six-dimensional representations, where the ``${\bf 6}$''  
is due to the three hypercolors and the two types of preons.
We then have for the fermions 
${\bf 6} \otimes {\bf 6} \otimes {\bf 6}= {\bf 20} \oplus {\bf 56} 
\oplus {\bf 70} \oplus {\bf 70}$, where only the 
lowest dimensional (antisymmetric) ${\bf 20}$ is needed to represent
the first generation of leptons, quarks, and hyperquarks and their 
antiparticles. The ${\bf 20}$ dimensional representation decomposes into 
4 hypercolor neutral leptons, 4 hypercolor neutral quarks, 
and 12 hyperquarks where the multiplicity of hypercolor (H)
is included as indicated below
\be
\label{SU(6)fermions}
\left ( 
\left(
\begin{array}{c} \nu \\ e^{-} \end{array} 
\begin{array}{c} \bar{\nu} \\ e^{+} \end{array} \right )
\left ( \begin{array}{c} u \\ d \end{array}
\begin{array}{c} \bar{u} \\ \bar{d} \end{array} \right )
\left(\begin{array}{c} \tilde{u} \\ \tilde{d} \end{array}
\begin{array}{c} \bar{\tilde{u}} \\ \bar{\tilde{d}} \end{array}\right)_{H}
\right )_P.
\ee
We mention that at energies where SU(6)$_P$ 
comes into play, anomaly freedom  of SU(6)$_P$ 
is guaranteed because by definition 
all its generators are left-right symmetric. Furthermore,  the charge sum 
of the above multiplet is zero.

\subsection{Preon intrinsic parity and weak interaction phenomenology}
\label{preonintrinsicparity}

In this subsection we wish to comment on the different intrinsic parity 
assignment for $T$ and $V$ preons (see Table~\ref{tab:preon-entities}) 
and its consequences for the effective preon bound state interactions
that have been discussed so far.

As shown in Table~\ref{tab:all fermions}, leptons and quarks have the same 
intrinsic parity whereas the two hyperquarks ${\tilde u}$ and ${\tilde d}$ 
have opposite intrinsic parities. Therefore, one can assign a weak isospin 
to leptons and quarks 
but not to hyperquarks. This means that only leptons and quarks can be 
divided into left- and right-handed sectors carrying different weak isospin 
$T_L$ and $T_R$.

At low energies below the partial unification scale $M_P$,
where chiral isospin symmetry is broken, 
this classification of leptons and quarks in separate 
chiral sectors having different weak isospin $T_L=1/2$ and $T_R=0$ 
goes hand in hand with left-right asymmetric weak interactions
mediated by the standard model gauge bosons $W_L$ and $Z$.
In contrast, hyperquarks do not possess a weak isospin and consequently 
do not participate in the standard model weak isospin interactions 
but only in left-right symmetric $\tilde{W}$, $N$, and $Z$ mediated 
interactions as discussed before.

For energies above $M_{P}$, hyperquark-quark transitions occur, and 
the left-right symmetric hyperquark interactions entail left-right 
symmetric weak interactions also for quarks and leptons coupling 
to the weak isospin $T_L$ and $T_R$ with equal strength because
at this scale the effective masses of the $W_L$, $W_R$, ${\tilde W}$ 
are of the same order as $M_P$. At this point the inner parity 
of preon bound states ceases to be a good quantum number.

Thus, preon intrinsic parities reveal themselves in 
different effective interactions of standard model fermions and of 
hyperquarks. At low energies of $10^{2}$ GeV, we have
left-right asymmetric interactions between particles
having definite intrinsic parities. At higher energies of $10^{9}$ GeV 
we have left-right symmetric interactions but intrinsic parity violating 
transitions involving hyperquarks and  neutrino-antineutrino pairs.

These aspects of the extended weak interaction  
have their origin in the different intrinsic parity assignments 
of the fundamental preon building blocks. Moreover, the existence of 
left-right symmetry at high and left-right asymmetry at low energies 
is seen to be closely connected with the production and decay of hyperquarks 
and hence to some extent explained in the preon model developed here.

\begin{table*}[htb]
\begin{center}
\begin{tabular}{ l  c  c  c }
\hline
gauge group & generators & gauge boson & energy scale \\
\hline
U(1)$_{Q}$ & 1 & $A_{Q}$ & $\Lambda_{Q} = 10^{-3} $ GeV \\
\hline
SU(3)$_{C}$ & 8 & $G_{C}$ & $\Lambda_{C} \cong 0.2 $ GeV \\
\hline
SU(2)$_{W_{L}} \times$ U(1)$_{Y}$  &  4 & $W^{-}_{L}; W^{+}_{L}; W^{0}_{L}; 
B^{0} $ 
& $M_{Z} = 91.18 $ GeV \\
\hline
SU(3)$_{H}$ &  8 & $ G_{H}$ & $\Lambda_{H} \cong 1700 $ GeV  \\
\hline 
SU(6)$_{P}$ & 35 &  $ G_{H}$  & \\
&  & $W^{-}_{L}; W^{+}_{L}; W^{0}_{L}; W^{-}_{R}; W^{+}_{R}; W^{0}_{R}; B^{0}$ 
& \\
&  & $\tilde{W}^{-}$; $\tilde{W}^{+}$ &  \\ 
& & $\left(N;\bar{N}\right)_{H}$ & $ M_{P} = 8.3 \times 10^{9}$ GeV\\
\hline
SU(9)$_{G}$ & 80 &  $ G_{P}$  & \\
& & $G_{C}$ & \\
& & $\left(X ; Y ; \bar{X}; \bar{Y}; U ;\bar{U}\right)_{C}$& \\
& & $\left(\tilde{X} ; \tilde{Y} ;\bar{\tilde{X}};\bar{\tilde{Y}};
\tilde{U}; \bar{\tilde{U}}\right)_{H}$& \\
& & $A_{Q} $& 
$ M_{G} = 1.2 \times 10^{16} $ GeV \\
\hline
\end{tabular}
\end{center}
\caption{\label{tab:gaugegroups}
Gauge groups, gauge bosons, and corresponding energy scales.}
\end{table*}

\section{Grand unification}
\label{sec:grandunification}

As motivated in the previous chapter we are now left with three gauge 
groups, namely SU(6)$_{P}$, SU(3)$_{C}$ and U(1)$_{Q}$. 
At the grand unification scale $M_{G}$,
the corresponding interactions have the same
strength and are described by a common coupling constant 
(see Fig.~\ref{figure:scattering}).
\begin{equation}
\label{G}
\alpha_{G} := \alpha_{P} = \alpha_{H} =
\alpha_{C} = \alpha_{Q}.
\end{equation}
At this scale the direct product, 
SU(6)$_{P}\times $SU(3)$_{C}\times$U(1)$_{Q}$ is
embedded in the larger gauge group
SU(9)$_{G} \supset \, \, $SU(6)$_{P}\times$SU(3)$_{C}\times$U(1)$_{Q}$.  
Thus, the unbroken fundamental gauge symmetry 
SU(3)$_{H}\times$SU(3)$_{C}\times$U(1)$_{Q}$ of
the preon model is also part of this larger group,  which finds
its expression in the equality of coupling constants according 
to Eqs.(\ref{G}). 

The grand unification group SU(9)$_{G}$ contains in addition to the
gauge bosons already included in  SU(6)$_{P}$ the
following color anti-triplet and hypercolor singlet gauge bosons generating 
transitions between quarks and leptons:
 
\begin{eqnarray}
\label{standardgut}
X & = &  \left(\begin{array}{c} TTV \\ TTV \end{array}\right); \quad \quad
Y = \left(\begin{array}{c} \bar{T}\bar{V}\bar{V} \\ \bar{T}\bar{V}\bar{V} 
\end{array}\right); \quad  \quad \nonumber\\
U & = & \left(\begin{array}{c} TTV \\ \bar{T}\bar{V}\bar{V} \end{array}\right) 
= \left( T\bar{V} \right),   
\end{eqnarray}
where the six-preon bound state $U$ reduces to a dipreon bound state
with respect to its quantum numbers. 
These bosons are responsible for proton decay, which has already been 
predicted by the SU(5) grand unification~\cite{geg74} 
as well as within the preon model~\cite{har82a}.

Furthermore, there are new hypercolor anti-triplet and color singlet gauge 
bosons generating transitions between hyperquarks and leptons 
\begin{eqnarray}
\tilde{X} & = & \left(\begin{array}{c} TT\bar{V} 
\\ TT\bar{V} \end{array}\right); 
\quad \quad 
\tilde{Y} = \left(\begin{array}{c} \bar{T}VV \\ \bar{T}VV \end{array}\right); 
\quad \quad \nonumber \\
\tilde{U} & = & \left(\begin{array}{c} TT\bar{V} \\ 
\bar{T}VV \end{array}\right) 
= \left( TV \right). 
\end{eqnarray}
Formally, the latter bosons are obtained from those in 
Eq.(\ref{standardgut}) by the hyperquark transformation.
The quantum numbers of the six-preon GUT bosons $X$ and $Y$ generating baryon 
decay into leptons and of the new GUT bosons $\tilde{X}$ and $\tilde{Y}$ 
mediating hyperbaryon decay into leptons are listed in 
Table~\ref{tab:GUTbosons}.
\begin{table}[htb]
\begin{center}
\begin{tabular}{l c c c c c }
\hline
state & content & ${\cal{P}}$ & $\Upsilon$ & $Q$ & $\Pi$ \\ 
\hline\hline
$X$ & $(4T,2V)$ &    
$ +2 $ & $ +\frac{2}{3} $ & $ +\frac{4}{3}$ & + \\
\hline
$\bar{X}$ & $(4\bar{T}, 2\bar{V})$   
& $ -2 $ & $ -\frac{2}{3} $ & $ -\frac{4}{3}$ & + \\
\hline
$Y$ &  $(2\bar{T}, 4\bar{V})$ &   
$ -2 $ & $ +\frac{2}{3} $ & $ -\frac{2}{3}$ & + \\ 
\hline
$\bar{Y}$ & $(2T, 4V)$   
& $ +2 $ & $ -\frac{2}{3} $ & $ +\frac{2}{3}$ & + \\
\hline
$\tilde{X}$ &  $(4T, 2\bar{V})$   
 & $ +\frac{2}{3} $ & $ +2 $ & $ +\frac{4}{3}$ & + \\
\hline
$\bar{\tilde{X}}$ &  $(4\bar{T}, 2V)$ 
& $ -\frac{2}{3} $ & $ -2 $ & $ -\frac{4}{3}$ & + \\
\hline
$\tilde{Y}$ &  $(2\bar{T}, 4V)$ &  
$ +\frac{2}{3} $ & $ -2 $ & $ -\frac{2}{3}$ & + \\
\hline
$\bar{\tilde{Y}}$ & $(2T, 4\bar{V})$   
& $ -\frac{2}{3} $ & $ +2 $ & $ +\frac{2}{3}$ & + \\
\hline
\end{tabular}  
\caption{\label{tab:GUTbosons}
Quantum numbers of the six-preon GUT gauge bosons $X$ and $Y$ 
responsible for baryon decay into leptons, and of the 
new GUT bosons $\tilde{X}$ and $\tilde{Y}$
mediating hyperbaryon decay into leptons.}
\end{center}
\end{table} 

\begin{figure}
\resizebox{0.4\textwidth}{!}{
\includegraphics{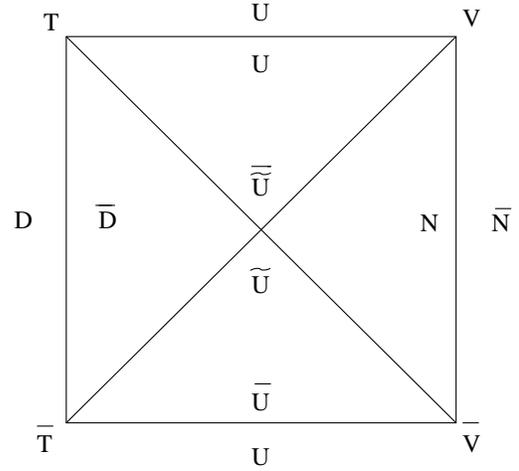}}
\caption{\label{figure:preonsquare} Preon square. 
The two fundamental preons $(T,V)$ and their antiparticles  
are placed at the corners of a square. The dipreon bound states  
and their antiparticles describing transitions between preons are 
placed along the edges $(N,U)$ and diagonals $(\tilde{U})$.
The dipreon states $D(TT)$ connecting $T$ and $\bar{T}$ can be represented
as a linear combination of $\tilde{U}$ and $U$ or $U$ and $N$. Thus
they are not included as independent gauge bosons in Table~\ref{tab:dipreons}.}
\end{figure}

The preon content of the dipreonic $U$ and of the neutralons $N$ 
already introduced in sect.~\ref{sec:partialunificationscheme} 
is given in Table~\ref{tab:dipreons}. The transitions
between the preons that are generated by these dipreonic bosons 
are graphically depicted in Fig.~\ref{figure:preonsquare}.
\begin{table}[htb]
\begin{center}
\begin{tabular}{l c c c c}
\hline
dipreon & ${\cal{P}}$ & $\Upsilon$ & $Q$ & $\Pi$ \\ 
\hline
$N\left(VV\right)$ & $ +\frac{2}{3} $ & $ -\frac{2}{3} $ & 0 & + \\ 
\cline{1-5}
$\bar{N}\left(\bar{V}\bar{V}\right) $ & $ -\frac{2}{3} $ & 
$ +\frac{2}{3} $ & 0 & + \\   
\hline\hline
$U\left(T\bar{V}\right) $ & $ 0 $ & $ +\frac{2}{3} $ & $ +\frac{1}{3}$ & +\\ 
\cline{1-5}
$\bar{U}\left(\bar{T}V\right) $ & $ 0 $ & $ -\frac{2}{3} $ & $ -\frac{1}{3}$ 
& + \\ 
\hline
$\tilde{U}\left(TV\right) $ & $ +\frac{2}{3} $ & $ 0 $ & $ +\frac{1}{3}$ & - \\ 
\cline{1-5}
$\bar{\tilde{U}}\left(\bar{T}\bar{V}\right)$ & $ -\frac{2}{3} $ & $ 0 $ & 
$ -\frac{1}{3} $ & - \\
\hline
\end{tabular}  
\caption{\label{tab:dipreons} Dipreon bound states and their quantum numbers.}
\end{center}
\end{table}

\subsection{Nucleon decay processes}
Nucleon decay into leptons is mediated by six preon bound states and 
in an analogous way by dipreon bound states. 
For example, in proton decay 
$p(uud) \longrightarrow e^{+} + \gamma$ the preons in 
the two $u$-quarks combine into an intermediate $X$-boson which 
then decays by rearrangement
into an ${\bar d}$ quark and an $e^{+}$   
\begin{eqnarray}
\label{proton_decay_X}
u \left(TTV\right) &+& u\left(TTV\right) 
\longrightarrow X \left(\begin{array}{c} TTV \\ 
TTV \end{array}\right) \nonumber \\
&\longrightarrow& \bar{d} \left(TVV\right) + e^{+}\left(TTT\right).
\end{eqnarray}
The remaining $d$-quark in the proton annihilates with the final state 
$\bar{d}$-quark into a photon.
An analogous proton decay process of a $(u\, d)$ quark pair in the proton 
involves the dipreon $U$ of Eq.(\ref{standardgut}) 
\begin{eqnarray}
\label{proton_decay_U}
u\left(TTV\right) &+& d \left(\bar{T}\bar{V}\bar{V}\right) \longrightarrow
U \left( T\bar{V} \right)  \nonumber \\
&\longrightarrow& \bar{u} \left(\bar{T}\bar{T}\bar{V}\right)+ 
e^{+} \left(TTT\right), 
\end{eqnarray}
where the final state is generated by creating two additional $T \bar{T}$ 
pairs from the vacuum. Note that the preon triality 
rule of Eq.(\ref{modulo3rule}) 
is broken at the grand unification scale.       

Similarly, neutron decay $n \longrightarrow \bar{\nu}+ \gamma$ 
proceeds either by rearrangement 
 \begin{eqnarray}
\label{neutron_decay_Y}
d \left(\bar{T}\bar{V}\bar{V}\right) &+& 
d \left(\bar{T}\bar{V}\bar{V}\right) \longrightarrow
Y \left(\begin{array}{c} \bar{T}\bar{V}\bar{V} \\ 
\bar{T}\bar{V}\bar{V} \end{array}\right) \nonumber \\
&\longrightarrow& \bar{u} \left(\bar{T}\bar{T}\bar{V}\right) +
\bar{\nu}\left(\bar{V}\bar{V}\bar{V}\right),
\end{eqnarray}
or via the dipreon $U$
\begin{eqnarray}
\label{neutron_decay_U}
u\left(TTV\right) &+& d \left(\bar{T}\bar{V}\bar{V}\right) \longrightarrow
U \left( T\bar{V} \right) \nonumber \\
&\longrightarrow& \bar{d}\left(TVV\right) + 
\bar{\nu}\left(\bar{V}\bar{V}\bar{V}\right)
\end{eqnarray}
requiring the creation of two $V{\bar V}$ pairs from the vacuum.
Thus, these nucleon decay processes can be described by six-preon bound states 
$X$ and $Y$, as well as by the dipreon bound state $U(T \bar{V})$.

In analogy, hypernucleon decay processes involve the corresponding 
$\tilde{X}$, $\tilde{Y}$, and $\tilde{U}(T V)$ bosons as discussed before
and are obtained from Eq.(\ref{proton_decay_X})-Eq.(\ref{neutron_decay_U}) 
by hyperquark transformation.
Common to these processes is that they cause a simultaneous 
violation of baryon- and lepton-numbers ($ \Delta B = \Delta L = - 1$), 
whereas ${\cal P}$ and ${\Upsilon}$
numbers are conserved. There are also $B-L$ violating proton 
decays~\cite{har82a} such as $p \to \nu + \pi^+$ 
which can be thought of as being generated from the $B-L$ conserving 
$p \to {\bar \nu} + \pi^+$ decay via a $\nu-{\bar \nu}$ oscillation 
as discussed in sect.~\ref{subsec:oscillation}.
 
For completeness we mention that between the color triplet six-preon bound 
states $X$, $Y$ and the dipreonic $U$ gauge bosons there are the 
following weak 
transition processes
\begin{eqnarray}
\label{ditosix}
X & \longrightarrow &  U + W^{+}; \quad
\quad Y \longrightarrow U + W^{-} \nonumber \\
\tilde{X} & \longrightarrow &  \tilde{U} + \tilde{W}^{+}; \quad \quad
\tilde{Y} \longrightarrow \tilde{U} + \tilde{W}^{-}.
\end{eqnarray}
The reactions in the second line are generated 
by applying the hyperquark transformation to the reactions in 
the first line, emphasizing once again that the hyperquark transformation 
is applicable to all preon bound states containing neutral $V$ preons. 

\subsection{Grand unification group SU(9)$_{G}$}
\label{sec:grandunificationgroup} 
After having discussed the additional bosons needed to enable transitions
from quarks and hyperquarks to leptons, we can now count the number of gauge
bosons involved. We find that the unitary group has to be as large as SU(9)
to accomodate all gauge bosons that have been introduced.  
The corresponding $9 \times 9$ matrix of gauge bosons      
is schematically shown below

\[ G_G=\left(
\begin{array}{c} \hspace{-1.8 cm} G_{C} \hspace{0.2 cm} 
\quad X\ Y\ U \hspace{0.7 cm}
 \quad \tilde{X}\ \tilde{Y}\ \tilde {U} \\ 
\hspace{-1.5 cm} \bar{X}\ \bar{Y}\ \bar{U} \quad G_{H} 
\quad \hspace{2.0 cm} N \\ 
\bar{\tilde{X}}\ \bar{\tilde{Y}}\ \bar{\tilde{U}} \quad 
\bar{N} \quad  
\left(\begin{array}{cc} W^{0}_{R} \qquad  W^{+}_{R}
\qquad  \tilde{W}^{+} \\
W^{-}_{R} \qquad  W^{0}_{L} \qquad  W^{+}_{L} \\  
\tilde{W}^{-} \qquad  W^{-}_{L} \qquad  B^{0} \end{array}\right)
\end{array}\right). \]
This scheme includes the $ 6 \times 6 $ matrix of the SU(6)$_{P}$ partial 
unification group $G_P$ of Eq.(\ref{pu_matrix}), 
the $3 \times 3$ matrix of the QCD gluons G$_C$ 
transforming according to SU(3)$_{C}$, 
four blocks of $3 \times 3$ matrices 
for the colored GUT bosons $X$, $Y$, and $U$, 
their antiparticles, as well as their hyperquark transformed states.  
In addition, we have to include the elementary photon associated 
with the group U(1)$_{Q}$. The generator $A_{Q}$ appears
as an admixture in the diagonal entries and must be included
in the counting leading to 80 generators in total. 

Representations for the fermionic preon bound states arise 
from the direct product of three fundamental
representations of dimension ${\bf 9}$, where the ${\bf 9}$ 
is due to the three color and three hypercolor degrees
of freedom associated with each preon. Note that the preon types $T$ and $V$ 
need not be included as a separate degree of freedom, i.e. we do not have an
$SU(18)$ because specifying the color and hypercolor representations
uniquely specifies the preon type (see Table~\ref{tab:preon-entities}).

We then have 
${\bf 9} \otimes {\bf 9} \otimes {\bf 9} = {\bf 84} \oplus {\bf 240} 
\oplus {\bf 240} \oplus {\bf 165}$, 
where only the lowest dimensional (antisymmetric) ${\bf 84}$ 
is needed to represent the three generations of leptons, quarks, 
and hyperquarks, as well as their antiparticles:
\be
\label{fermions & preons}
\left(\left(
\begin{array}{c} \nu \\ e^{-} \end{array} 
\begin{array}{c} \bar{\nu} \\ e^{+} \end{array} \right ) \, 
\left(\begin{array}{c} u \\ d \end{array}
\begin{array}{c} \bar{u} \\ \bar{d} \end{array}\right)_{C}
\left(\begin{array}{c} \tilde{u} \\ \tilde{d} \end{array}
\begin{array}{c} \bar{\tilde{u}} \\ \bar{\tilde{d}} \end{array}\right)_{H}
\right)_{G}.
\ee
The ${\bf 84}$ dimensional fermion representation decomposes as follows:
12 leptons, 36 quarks, and 36 hyperquarks, 
where the multiplicities of color (C),
hypercolor (H) and generation number (G) for the bound states are included.

\vspace{0.5 cm}
\section{Running couplings and energy scales}
\label{sec:runningcoupling}
On the basis of the model described so far, 
we discuss in this section the momentum dependence 
of the couplings $\alpha(q^2)$ described in 
sect.~\ref{sec:partialunificationscheme}, 
where $q^2$ is the momentum transfer exchanged 
in the interaction. Our aim is to determine the scale,
where the lightest hyperquark bound states appear,
that is the mass scale $\Lambda_{H}$ of hyperchromodynamics, where
$\alpha_{H}$ is of order ${\cal O}(1)$. 
This is analogous to QCD where the lightest
quark bound states (pions) appear at $\Lambda_{C}$, 
i.e. the scale where $\alpha_{C}$ is of order ${\cal O}(1)$. 

In analogy to QCD, $\alpha_{H}(q^{2})$ will decrease 
with increasing $q^{2}$ and the slope $b_{H}$ 
of this decrease depends on the number of hyperquark flavors along the way 
to higher energies in the same manner as the slope $b_{C}$ 
of $\alpha_{C}(q^{2})$ depends on the number of quark flavors appearing with 
increasing $q^{2}$. It should be clear that we cannot
determine the exact position of each single hyperquark flavor but only
an average value of these hyperquark masses denoted as $m_{hq}$.

To fix the high $q^{2}$ end of the running couplings we use the 
constraints defining partial and grand unification of 
Eq.(\ref{PUH}) and Eq.(\ref{G}) respectively as detailed in the next section, 
where we also study how the $\beta$-functions of the different 
effective gauge groups change when $q^2$ crosses the hyperquark mass scale.

\subsection{$\beta$-functions of the SU(N) and U(1) gauge groups}
In non-abelian gauge theories the slope of the 
running couplings is determined to first order by the $\beta$-function 
$b_{i}$  
\begin{equation}
\label{betaf}
b_{i} = \frac{11}{3} N_{i} - \frac{2}{3} n_{f}.
\end{equation}
where the index $i$ stands for the different gauge groups, 
$N_{i}$ is number of degrees of freedom of the $SU(N_i)$ group,
and $n_{f}$ is the number of fermion flavors. 
For the running coupling one has in one-loop 
approximation~\cite{grw73}  
\begin{equation}
\label{alphaq}
\frac{1}{\alpha_{i}(q^2)} = \frac{1}{\alpha_{i}(\Lambda_{i}^{2})} + 
\frac{b_{i}}{4\pi} 
\ln\left(\frac{q^{2}}{\Lambda_{i}^{2}}\right).
\end{equation}
For the present purposes higher order approximations of $\alpha_{i}$ 
can be neglected. 
Eq.(\ref{alphaq}) connects an a priori unknown high energy scale 
$q^2=M^2$ to a known low-energy scale $\Lambda^2$ such as 
for example, $\Lambda_{C}$ for QCD. Constraints for the high energy scale 
are obtained from the equality of certain coupling constants at this 
mass scale.
\begin{widetext}
\begin{figure*}[htb] 
\resizebox{0.8\textwidth}{!}{\includegraphics{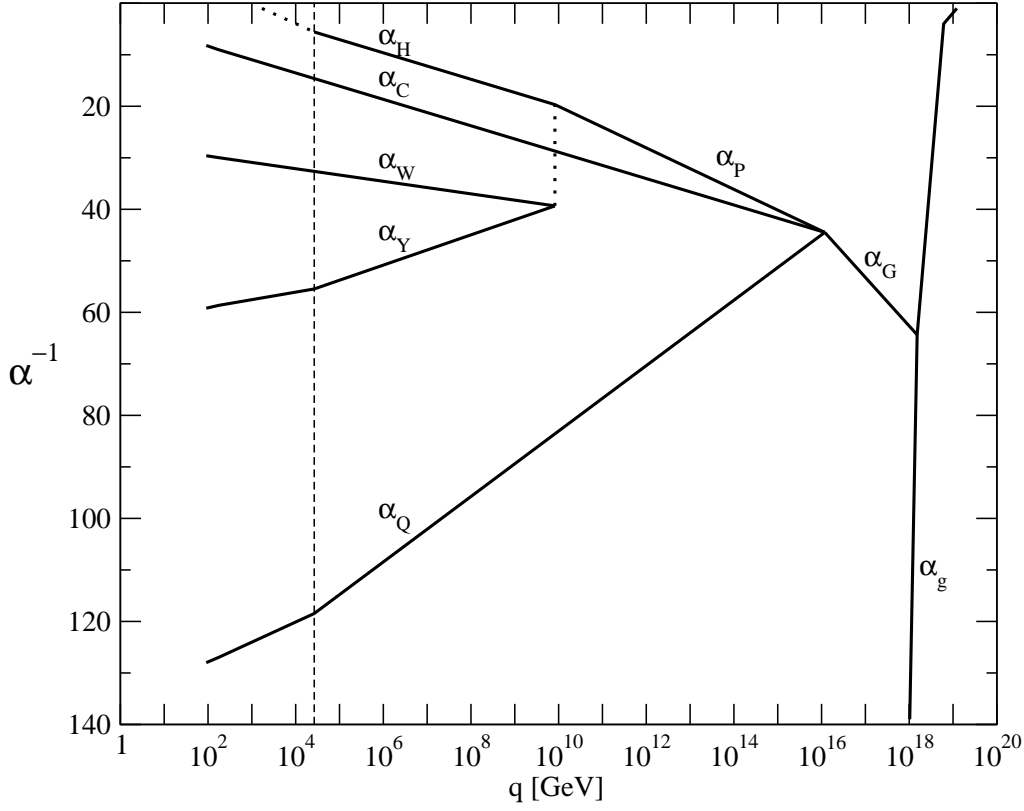}}
\vspace{0.5cm}
\caption{\label{figure:scattering} 
Running coupling constants. The 
dashed vertical line indicates the expected hyperquark mass scale.
Because hyperquarks are hypercolor triplets and charged weak isosinglets, 
they only affect the $SU(3)_{H}$ coupling $\alpha_{H}$ and 
the  $U(1)$ gauge couplings $\alpha_{Q}$ and $\alpha_{Y}$.
The dotted vertical line at $10^{9}$ GeV 
indicates the partial unification scale 
where the weak isospin and hypercharge couplings converge
and are unified with the hypercolor interaction to a 
common coupling $\alpha_{P}$. The three gauge couplings 
$\alpha_{P}$, $\alpha_{C}$, and $\alpha_{Q}$ meet at the grand 
unification scale of $10^{16}$ GeV.
Finally, from the grand unification scale to the point 
where $\alpha_{G}$ meets the gravitational coupling
$\alpha_{g}$ $(\cong 10^{18}$ GeV), preons behave as quasifree particles.}
\end{figure*}
\end{widetext}
\subsubsection{The $\beta$-functions of SU(3)$_{C}$ and SU(3)$_{H}$}
For the non-abelian groups SU(3)$_{C}$ and SU(3)$_{H}$ and 
where $N_{C}=N_{H}=3$ we have
\begin{eqnarray}
\label{betavalues}
b_{C} &=& \frac{11}{3}\, N_{C} - \frac{2}{3} \,
\left( n_{d} + n_{u} \right) \nonumber \\ 
b_{H} &=& \frac{11}{3}\, N_{H} - \frac{2}{3} \, 
\left( n_{\tilde{d}} + n_{\tilde{u}} \right).
\end{eqnarray}
Because there are three generations for each fermionic bound 
state type~\cite{abms05} 
\begin{equation}
n_{d} = n_{u} = n_{\tilde{d}} =  n_{\tilde{u}} = 3, 
\end{equation}
we obtain for both gauge groups the same $b$ value
\begin{equation} 
b_{C} = b_{H}=7.
\end{equation}

\subsubsection{The $\beta$-function of U(1)$_{Q}$}
The running coupling constant of QED reads
\begin{equation}
\label{alpha_{Q}}
\frac{1}{\alpha_{Q}(q^2)} = \frac{1}{\alpha_{Q}(\Lambda_{Q})} + 
\frac{b_{Q}}{4\pi} 
\ln\left(\frac{q^{2}}{\Lambda_{Q}^{2}}\right).
\end{equation}
Here, $b_{Q}$ is defined as 
\begin{equation}
b_{Q} = - \frac{2}{3} \sum_{i} Q_{i}^{2},
\end{equation}
where $Q_{i}^{2}$ is the square 
of the fermion charge (see Table~\ref{tab:isospin}),
and the sum extends over all fermions. 
Thus, the $\beta$-function of QED is 
\begin{eqnarray}
b_{Q} &=& - \frac{2}{3} \, 2 \, \Biggl( n_{e} + 3 \left(\frac{1}{9} n_{d} 
+ \frac{4}{9} n_{u} \right) \nonumber \\ 
& & + 3 \left( \frac{1}{9} n_{\tilde{d}} + \frac{4}{9} 
n_{\tilde{u}}\right)\Biggr ),
\end{eqnarray}
where $n_{e}=3$ is the number of leptonic flavors. 
The identical contributions of the left- and right handed fermion
charges are taken into account by an overall factor 2,
and the color and hypercolor multiplicities are indicated by factors 3
multiplying the quark and hyperquark contributions.

At the hyperquark scale $m_{hq}$ the $\beta$-function
$b_{Q}$ changes 
\begin{eqnarray}
b_{Q} &=& - \frac{32}{3} \ {\rm (without\  hyperquarks)}, \nonumber \\  
b_{Q} &=& - \frac{52}{3} \ {\rm (with \ hyperquarks)}
\end{eqnarray}
due to the additional contribution of the hyperquarks.

\subsubsection{Electroweak $\beta$-functions for U(1)$_Y$ and SU(2)$_{W}$}
For the U(1)$_{Y}$ group we replace in Eq.(\ref{alpha_{Q}}) 
$\alpha_{Q}$ by $\alpha_{Y}$ and $b_{Q}$ by $b_{Y}$ which
is defined as 
\begin{equation}
b_{Y} = - \frac{2}{5} \sum_{i} Y_{i}^{2},
\end{equation}
where the sum extends over the fermionic preon bound states.
In addition, we have to take into account that left- and right-handed fermions
give different contributions. 
As noted before hyperquarks are weak isospin singlets ($ T_{3}= 0$) 
and therefore have a left-right symmetric coupling, whereas 
quarks and leptons couple left-right asymmetrically.

We then obtain for the $\beta$-function according to
Table~\ref{tab:isospin} 
\begin{eqnarray}
b_{Y} & = & - \frac{2}{5}  \Biggl(\frac{1}{4} n_{\nu} + \frac{5}{4} n_{e} 
 + 3 \left( \frac{5}{36} n_{d} + \frac{17}{36} n_{u} \right) \nonumber \\ 
& & + 3  \left(\frac{2}{9} n_{\tilde{d}} + 
\frac{8}{9} n_{\tilde{u}}\right)\Biggr ),
\end{eqnarray}
with $n_{\nu}=3$ and where the factors in front of the $n_i$ come
from adding left and right handed contributions in 
Eq.(\ref{upsilonsum_{hq}}). Note that $b_{Y}$ changes at the hyperquark 
mass scale $m_{hq}$ due to the contribution of the hyperquarks as 
\begin{eqnarray}
b_{Y} &=& - 4  \ {\rm (without \  hyperquarks)}, \nonumber \\
b_{Y} &=& - 8  \ {\rm (with \  hyperquarks)}.
\end{eqnarray}

For the non-abelian weak isospin group SU(2)$_{W}$ the corresponding
$\beta$-function reads
\begin{equation}
b_{W} = \frac{11}{3} N \, - \frac{1}{3} \, 
\left( n_{\nu} + n_{e} + n_{d} + n_{u} \right) \,
\end{equation}
where $N=2$. 
Here, only left-handed leptons and quarks contribute and we obtain 
the standard model value $\quad b_{W}=\frac{10}{3}$.

\subsubsection{The $\beta$-function of the partial unification group SU(6)} 
Here, we deal with the nonabelian gauge group $SU(6)$ as
discussed in sect.~\ref{sec:partialunificationscheme}.
At the $M_{P}$ scale the Majorana description of the neutrinos 
implies that the right-handed neutrinos are identical with antineutrinos,
i.e., $ \nu_{R} \equiv \bar{\nu}$.  Therefore, only the left-handed neutrinos 
are included in the summation over the left- and right-handed fermions 
\begin{equation}
\label{bPU}
b_{P} = \frac{11}{3}\, N_{P} - \frac{1}{3} \, 
\left(n_{\nu} + 2 \, \left(n_{e} + n_{d} + n_{u} + n_{\tilde{d}} + 
n_{\tilde{u}} \right)\right),
\end{equation}
which gives $b_{P}=11$ for $N_P=6$.

We point out that two-state Majorana neutrinos provide the only
consistent description in the present framework.
A four-state Dirac neutrino would reduce the $\beta$-function to $b_{P} = 10$. 
As will be seen in section~\ref{subsec:energyscales}, 
the latter would lead to inacceptable 
results for the energy scales $M_{G}$, $M_{P}$, and
$m_{hq}$. In particular, it would imply 
a hyperquark mass scale below the masses of W and Z bosons,
and a much too short proton lifetime, both of which are in 
contradiction to experimental facts.  

\subsubsection{The $\beta$-function of the grand unification group SU(9)}
At this scale the different bosons can be accommodated into a larger 
gauge group $SU(9)$. By an analogous counting of the fermions as in 
Eq.(\ref{bPU}) we obtain
\begin{equation}
b_{G} = \frac{11}{3}\, N_{G} - \frac{1}{3} \, 
\left(n_{\nu} + 2 \, \left(n_{e} + n_{d} + n_{u} + n_{\tilde{d}} + 
n_{\tilde{u}} \right)\right) 
\end{equation}
which gives $\quad b_{G} = 22$ for $N_G=9$.

\subsection{Calculation of energy scales}
\label{subsec:energyscales}
To obtain numerical values for $M_{G}$, $M_{P}$ and $m_{hq}$ one starts
from the following constraints.
First, from Eq. (\ref{EQWY}) we have at $M_{P}$ 
\begin{eqnarray}
\label{cond1}
\frac{1}{\alpha_{W}(M_{P}^{2})} &=& \frac{1}{\alpha_{Y}(M_{P}^{2})} 
\nonumber \\
 & & \nonumber \\ 
\frac{1}{\alpha_{W}(m_{t}^{2})} + \frac{b_{W}}{4\pi} 
\ln\left(\frac{M_{P}^{2}}{m_{t}^{2}}\right) &=& 
\frac{1}{\alpha_{Y}(m_{t}^{2})} \nonumber \\  
+ \frac{b_{Yt}}{4\pi} 
\ln\left(\frac{m_{hq}^{2}}{m_{t}^{2}}\right)
+ \frac{b_{Y}}{4\pi} \!\!\!\!\!\! &\!\! & \!\!  
\!\!\!\ln\left(\frac{M_{P}^{2}}{m_{hq}^{2}}\right), 
\end{eqnarray}
where the evolution starts at the top quark mass $m_{t}$.
According to Eq.(\ref{alphaq}) the first two terms on the right-hand side 
can be written as 
$1/\alpha_{Y}(m_{hq})$, and the third term on the right-hand side 
describes the evolution from $m_{hq}$, at which point we have
to include the hyperquarks to the partial unification scale 
$M_{P}$.

Second, from Eq.(\ref{PUW}) and Eq.(\ref{G}) evaluated at $M_{G}$ follows
\begin{eqnarray}
\label{cond2}
\frac{1}{\alpha_{C}(M_{G}^{2})} &=& \frac{1}{\alpha_{P}(M_{G}^{2})} 
\nonumber \\
 & & \nonumber \\ 
\frac{1}{\alpha_{C}(m_{t}^{2})} &+& \frac{b_{C}}{4\pi} 
\ln\left(\frac{M_{G}^{2}}{m_{t}^{2}}\right) \nonumber \\ 
&=& \frac{1}{2} \left(\frac{1}{\alpha_{W}(m_{t}^{2})}
+ \frac{b_{W}}{4\pi} 
\ln\left(\frac{M_{P}^{2}}{m_{t}^{2}}\right)\right) \nonumber \\  
&+& \frac{b_{P}}{4\pi} \ln\left(\frac{M_{G}^{2}}{M_{P}^{2}}\right),
\end{eqnarray}
where the factor 2 in the denominator of the first term on the 
right-hand side comes from the strength doubling of $\alpha_W$ at
the partial unification scale according to Eq.(\ref{PUH}).

Third, from Eq.(\ref{G}) also follows
\begin{eqnarray}
\label{cond3}
\frac{1}{\alpha_{C}(M_{G}^{2})} &=& \frac{1}{\alpha_{Q}(M_{G}^{2})} 
\nonumber \\
 & & \nonumber \\ 
\frac{1}{\alpha_{C}(m_{t}^{2})} + \frac{b_{C}}{4\pi} 
\ln\left(\frac{M_{G}^{2}}{m_{t}^{2}}\right) &=& 
\frac{1}{\alpha_{Q}(m_{t}^{2})}  \nonumber \\ 
+ \frac{b_{Q}}{4\pi} 
\ln\left(\frac{m_{hq}^{2}}{m_{t}^{2}}\right) \!\!\! 
 &+& \!\!\!\frac{b_{Q}}{4\pi} 
\ln\left(\frac{M_{G}^{2}}{m_{hq}^{2}}\right).
\end{eqnarray}

The energy scales where these conditions hold can now
be determined by solving the above three equations (\ref{cond1}-\ref{cond3}) 
with three unknowns. This gives
\begin{eqnarray}
\label{massscales}
m_{hq} &=&  2.6 \times 10^{4} \,\, \mbox{GeV} \nonumber \\
M_{P}  &=&  8.3 \times 10^{9}  \,\, \mbox{GeV} \nonumber \\
M_{G}  &=&  1.2 \times 10^{16} \,\,  \mbox{GeV.} 
\end{eqnarray}

In order to calculate the low energy scale $\Lambda_{H}$ we 
use Eq.(\ref{alphaq}) 
and start the evolution at the partial unification scale $M_{P}$ 
where according to Eq.(\ref{PUH}) the following 
equality of coupling constants holds 
$\alpha_{P} = \alpha_{H}= 2\alpha_{W} $. By backextrapolation we obtain
\begin{eqnarray}
\label{lambda}
\frac{1}{\alpha_{H}(\Lambda_{H}^2)} &=& \frac{1}{\alpha_{H}(M_{P}^{2})}
-\frac{b_{H}}{4\pi} \ln\left(\frac{M_{P}^{2}}{m_{hq}^{2}}\right)  \nonumber \\
&-&\frac{b_{H1}}{4\pi} \ln\left(\frac{m_{hq}^{2}}{\Lambda_{H}^{2}}\right),
\end{eqnarray}
where $b_{H1} = \frac{31}{3}$ is calculated according to 
Eq.(\ref{betavalues}) assuming $n_{\tilde{d}}=1$ and
$n_{\tilde{u}}=0$ for the contribution of the lightest hyperquark 
below the average scale $m_{hq}$.
The slope $b_{H1}$ corresponds 
to the dotted line segment of $\alpha_H$ in 
the left upper corner of Fig.~\ref{figure:scattering}. 
We then demand $\alpha_{H}(\Lambda_{H}^{2}) \cong {\cal O}(1)$ which,
using the numerical values in Eq.(\ref{massscales}), leads to a  
prediction of the infrared cut off mass of hypercolor interactions
\begin{equation}
\Lambda_{H} \cong 1700 \, \mbox{GeV.}
\end{equation}
This is different from the model of Harari-Seiberg which gives
a $\Lambda_H$ of order $10^9$ GeV.
The numerical values of the $\beta$-functions, coupling constants,
and energy scales calculated in this section 
are compiled in Table~\ref{tab:numerical}.

With the constraint $m_{hq1} \geq \Lambda_{H}$ where $m_{hq1}$ 
denotes the mass of the lightest hyperquark, 
we obtain for the weak decay lifetime of hyperquark bound states, e.g.,
a hyperpion
\begin{equation} 
\label{LT}
\tau_{{\tilde \pi}} \cong \frac{1}{\alpha_{P}^{2}} 
\frac{M_{P}^{4}}{m_{hq1}^{5}}
\leq 100 \, s.
\end{equation}
For the neutrino masses according we obtain according to 
Eq.(\ref{seesaw})
\begin{eqnarray}       
m_{\nu_{e}} &=&  3.3 \times 10^{-8} \,\, \mbox{eV},  \nonumber \\ 
m_{\nu_{\mu}} &=&  1.3 \times 10^{-3} \,\, \mbox{eV},  \nonumber \\ 
m_{\nu_{\tau}} &=& 3.8 \times 10^{-1} \,\, \mbox{eV}   
\end{eqnarray}  
compared to $m_{\nu_{\mu}} =  8.8 \times 10^{-3}$ eV and 
$m_{\nu_{\tau}} = 5.0 \times 10^{-2}$ eV obtained  
from upper limits of experimental neutrino squared mass 
differences~\cite{gon09}. 

Furthermore, the large value of the SU(9)$_{G}$ boson masses 
$ M_{G} =  1.2 \times 10^{16}$ GeV results in the following value for 
the proton lifetime
\begin{equation} 
\tau_{p} \cong \frac{1}{\alpha_{G}^{2}} \frac{M_{G}^{4}}{m_{p}^{5}} 
\cong 10^{35} \ \rm{y},
\end{equation}
where $m_{p}=938.3$ MeV is the proton mass. 
This is in good agreement with the experimental
lower limit $ \tau_{p} > 5.5 \times 10^{33}$ y~\cite{imb90}.

\begin{center}
\begin{table*}
\begin{tabular}{ l c c c c c }
& &  M$_{Z} = 91.18 \pm 0.02$ GeV  & & &  \\
\hline
SU(3)$_{C}$ & SU(2)$_{W}$ & U(1)$_{Y}$ & U(1)$_{Q}$ \\
$b_{Cz} = \frac{23}{3}$ & $b_{Wz} = \frac{11}{3}$ & 
$b_{Yz} =-\frac{103}{30}$ & $b_{Qz} =-\frac{80}{9}$ \\ 
$\alpha_{Cz}^{-1} = 8.24(12)$ & $\alpha_{Wz}^{-1} = 29.57(05)$ &
$\alpha_{Yz}^{-1} = 59.00(04)$ &$\alpha_{Qz}^{-1} = 127.90(02)$ \\
\hline
& & & & \\
\end{tabular}
\begin{tabular}{ l c c c c }
& & m$_{t} = 176.9 \pm 4.0$ GeV  & & \\
\hline
$b_{C} = 7 $ & $b_{W} = \frac{10}{3}$ & $b_{Yt} = - 4 $ & $ b_{Qt} = 
-\frac{32}{3}$ \\ 
$\alpha_{Ct}^{-1} = 9.05(15) $ & $\quad \alpha_{Wt}^{-1} = 
29.96(06) $ &
$\quad \alpha_{Yt}^{-1} = 58.64(05)$ &$\quad \alpha_{Qt}^{-1} = 
126.96(05) $ \\
\hline
& & & & \\
\end{tabular}
\begin{tabular}{ l c c c c c }
&  $m_{hq} = \left(26.3 \pm 2.3 \right)10^{3}$ GeV & 
($\Lambda_{H} = \left( 1.66 \pm 0.34 \right)10^{3}$ GeV) &  & &   \\
\hline
SU(3)$_{H}$ & SU(3)$_{C}$ & SU(2)$_{W}$ & U(1)$_{Y}$ & U(1)$_{Q}$ \\
$b_{H} = 7 $ & $b_{C} = 7 $ & $b_{W} = \frac{10}{3}$ & 
$b_{Y} = - 8 $ & $ b_{Q} = -\frac{52}{3}$ \\ 
$\alpha_{Hh}^{-1} = 5.58(20) $ & $\alpha_{Ch}^{-1} = 14.63(18) $ &
$\alpha_{Wh}^{-1} = 32.62(08) $ & $\alpha_{Yh}^{-1} = 55.45(07) $ 
&$\alpha_{Qh}^{-1} = 118.47(09) $ \\
\hline
& & & & \\
\end{tabular}
\begin{tabular}{ l c c  }
& $M_{P} = \left(8.31 \pm 0.88 \right)10^{9}$ GeV  & \\ 
\hline
SU(6)$_{P}$ & SU(3)$_{C}$ & U(1)$_{Q}$  \\
$b_{P} = 11 $ & $ b_{C} = 7 $ & $ b_{Q} = -\frac{52}{3}$ \\ 
$\alpha_{P}^{-1} = 19.66(08) $ & $\alpha_{Cp}^{-1} = 28.73(18) $ &
$\alpha_{Qp}^{-1} = 83.57(09) $ \\
\hline
& &  \\
& $M_{G} = \left(1.17 \pm 0.12 \right)10^{16}$ GeV &  \\
\hline
& SU(9)$_{G}$  &  \\
& $b_{G} = 22$ &  \\ 
& $\alpha_{G}^{-1} = 44.48(27) $ &  \\
\hline \\
& $M_{pr} = \left(1.56 \pm 0.01 \right)10^{18}$ GeV & \\
\hline 
& $\alpha_{g}^{-1} = 61.20(37) $  & \\
\end{tabular}
\caption{\label{tab:numerical}
Numerical values of $\beta$ functions and coupling 
constants at the corresponding energy scales. The 
indicated errors are due to the experimental input values. }
\end{table*}
\end{center}

We close this section with a remark on the evolution of $\alpha_G$ 
for still higher momentum transfers.
The grand unification scale $M_{G}$ calculated here 
is near the Planck scale of quantum gravity $M_{Pl}$ given by
\begin{equation}
M_{Pl}^{2} = \frac{\hbar c}{\gamma_{G}} , \qquad 
\quad \alpha_{g} = \frac{q^2}{M_{Pl}^{2}},
\end{equation}
where $\gamma_{G}$ is the Newton gravitational constant.
The coupling $\alpha_{g}$ is the only nonrenormalized coupling, 
having a linear dependence on the momentum transfer $q^2$. 

Explicit preon degrees of freedom will become important at 
the fundamental preon scale defined by the constraint 
$ \alpha_{G}(M_{pr}^2) = \alpha_{g}(M_{pr}^2)$.  
We can write
\begin{equation}
\frac{1}{\alpha_{G}(M_{G}^2)} + \frac{b_{G}}{4\pi} 
\ln\left(\frac{M_{pr}^{2}}{M_{G}^{2}}\right) = 
\frac{M_{Pl}^{2}}{M_{pr}^{2}}.
\end{equation}
From this constraint we obtain using $ M_{Pl} = 1.22 \times 10^{19}$ GeV 
for the Planck scale 
\begin{equation}
M_{pr} = 1.56 \times 10^{18} \, \mbox{GeV} 
\end{equation}
for the scale $M_{pr}$ where an asymptotically free preon dynamics 
is expected.

\subsection{Hyperquark and hyperhadron mass scales} 

We are now in the position to make statements about the energy scale 
where the hyperquarks and their bound states presumably occur.  
As mentioned before, 
it is reasonable to assume that the new particles  do not all appear at the 
same threshold level but that their mass values are distributed 
over a certain range. Thus, the present result $m_{hq}=26$ TeV
provides an average energy level but gives no information concerning 
the size of the mass range over which the particles are distributed.
However, because of the missing isospin 
degree of freedom the spectrum of hyperhadrons 
is expected to be much wider than that of ordinary hadrons. 

For quark bound states we have the result $ m_{\pi} \sim \Lambda_{C}$. 
Analogously, for the bound states of hyperquarks we have 
$ m_{\tilde{\pi}} \sim \Lambda_{H}$. This leads 
to the following estimate for the hyperhadron mass
\begin{equation}
\label{hybs}
m_{\tilde{h}} \cong \frac{\Lambda_{H}}{\Lambda_{C}}\, m_{h}.
\end{equation}
With the numerical values $ m_{\pi} = 140$ MeV, $\Lambda_{C}= 100$ MeV,
and $\Lambda_{H}= 1700$ GeV, we arrive at a hyperpion mass 
$ m_{\tilde{\pi}} = 1100 $ GeV. Similarly, using the proton mass 
$m_{p}=938$ MeV as input we find hyperbaryon masses 
to be of order $ m_{\tilde{B}} \cong 10^4$ GeV. 
These results are compiled in Table~\ref{tab:lowhypermeson}.

We conclude this section with some remarks pertaining to the 
hyperquark and hyperhadron masses and lifetimes.
Because hyperquarks $\tilde{d}$ and $\tilde{u}$ have different
intrinsic parities they do not form strong 
isospin doublets. Thus, unlike the quarks of the first generation 
their masses need not be close to each other. 
For the lightest hyperquark ($\tilde{d}$) we find a constituent mass of order 
$m_{\tilde{d}} \cong \Lambda_{H} \cong 1700$ GeV. 
The mass of the hyperquark ($\tilde{u}$) could be substantially 
higher, and is conjectured to be of order  
$\left(m_{\tilde{d}} + \Lambda_{H}\right)\cong 3400$ GeV. 
Consequently, unlike the standard model pions, 
hyperpions do not form an isospin triplet and we expect that  
charged hyperpions are heavier than neutral ones.
Furthermore, because of the missing isospin degree of freedom, 
the lowest-lying hyperhadrons have a symmetric 
spin configuration corresponding to spin $\frac{3}{2}$ 
for hyperbaryons and spin $1$ for hypermesons.

While neutral hyperquark bound states can decay electromagnetically,
the charged ones can only decay via $\tilde{W}$ emission as discussed
in sect.~\ref{subsec:hypermeson}.  
Therefore, the lifetimes of the charged hyperquark bound states are
according to Eq.(\ref{LT}) with $\tau_{hq} \leq 100$ s substantially 
longer than the lifetimes of the neutral ones.
As to the experimental signature, a neutral hyperpion  
decays electromagnetically into two photons with lifetime 
$\tau_{\tilde\pi}= \hbar/\Gamma_{\tilde \pi} \leq 10^{-27}$ s
if one assumes a decay width $\Gamma_{\tilde \pi}= 800$ GeV
derived from $\Gamma_{\tilde{h}} 
\cong \frac{\Lambda_{H}}{\Lambda_{C}}\, \Gamma_{h}$ in analogy 
to Eq.(\ref{hybs}). 
\begin{figure}
\resizebox{0.3\textwidth}{!}{
\includegraphics{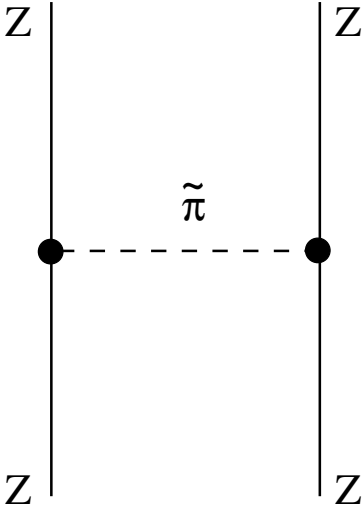}}
\caption{\label{figure:h0boson} 
$ZZ$ bound state $H^0$ generated via ${\tilde{\pi}}$ exchange.}
\end{figure}

Hypermeson masses of order TeV correspond to a short range 
Yukawa coupling interaction with range 
$\hbar/m_{\tilde{\pi}}c^{2}\cong 10^{-19}$ m.
At energies of several tens of TeV, hyperquarks couple only to the Z boson 
and the photon. As a result, there is 
a coupling of the neutral hyperpion to the Z boson.
This short-range interaction could lead to the formation of a scalar ZZ  
bound state with spin 0, called  $H^{0}$, as shown in 
Fig.~\ref{figure:h0boson}, which could be a viable 
candidate for solving the unitarity problem in  
W-W-scattering~\cite{lqt77}.
For the H$^0$ boson mass and decay width an estimate 
according to Ref.~\cite{lqt77} using $\lambda = G_{f}m_{H^{0}}^{2}/\sqrt{2}$,
where $ G_{f}^{-1/2}\cong 293$ GeV 
is the Fermi constant and $1\leq \lambda \leq 2$
gives $m_{H^{0}} \cong 350 - 500$ GeV and $\Gamma_{H^{0}} \cong 40$ GeV
corresponding to the lifetime 
$\tau_{H^{0}}= \hbar/\Gamma_{H^{0}} \cong 10^{-26}$ s.
\begin{table}[htb]
\begin{center}
\begin{tabular}{l c c c c c}
\hline\hline
boson & spin & $\Pi$ & mass [GeV] & lifetime [s] & decay-mode \\ 
\hline
$H^{0}$ & 0 & + & $ 
\cong 350 - 500$ & $10^{-26}$ & $Z$, $W^{+}W^{-}$,$f\bar{f}$ 
\\ 
\hline
$\tilde{\pi}^{0}$ & 1 & - & $ \cong 1100 $ & $10^{-27}$ & $H^{0}$,$Z$, 
$\gamma$ \\
\hline  
$\tilde{\pi}^{\pm}$ & 1 & + & $ \approx 3000 $ & $\leq 100 $ &$f\bar{f}$ \\ 
\hline\hline 
$\tilde{\Delta}^{-}$ & $\frac{3}{2}$ & + & $ \approx 10^{4} $ & $\leq 1 $ & 
$\Delta^{-}$ \\
\hline
$\tilde{\Delta}^{++}$ & $\frac{3}{2}$ & - & $ \approx 10^{4} $ & $\leq 1 $ & 
$\Delta^{++}$ \\
\hline\hline
\end{tabular}  
\caption{The H$^0$ boson and low lying hyperhadrons}
\label{tab:lowhypermeson}
\end{center}
\end{table} 

\section{Summary and outlook}
\label{sec:summary}

The standard model leaves many questions unanswered; for example, 
why leptons and quarks share the same weak interaction, 
mediated by heavy vector bosons coupling differently to left- and 
right-handed quarks and leptons. This fact, among others, points to a deeper 
connection between leptons and quarks. In the preon model, weak interactions 
are qualitatively understood as a residual force that is associated with the 
preon number of the bound state. On the other hand,  
a model which contains only the unbroken gauge interactions 
SU(3)$_{H} \times $SU(3)$_{C} \times $ U(1)$_{Q}$ between preons
does not readily lend itself to a quantitative description of the left-right 
asymmetric weak interactions between preon bound states at low energies.
Furthermore, it does not provide a dynamical explanation for certain new 
phenomena predicted by the present theory, 
for example, the decay of hyperquarks into quarks. 
Therefore, we have attempted to make some progress by considering 
approximate effective gauge theories on the level of preon bound states.

In particular, we have investigated the bosonic sector of the preon model 
in some detail. We have shown that the introduction of hyperquarks 
in Harari's theory requires new classes of effective gauge bosons, 
called $\tilde{W}$ and  $N$ in order to describe weak 
transitions among hyperquarks and between hyperquarks and quarks. 
The presence of these additional gauge bosons leads to an extension 
of the standard SU(2)$_{W_L}\times$U(1)$_{Y}$ electroweak theory to
a larger gauge group emerging at an energy $M_{P} \cong 10^{9}$ GeV. 
At this scale, 9 left-right symmetric weak bosons, 
8 hypergluons, and 18 neutralons provide the 35 generators of 
an effective SU(6)$_{P}$ gauge group, refered to as partial unification group.
This scheme predicts a Weinberg angle $\sin^2 \Theta_{W} = 6/13$
at $M_{P} \cong 10^{9}$ GeV, which is twice its experimental value at 
$M_{Z}=91.2$ GeV. Furthermore, it shows that the breaking of left-right 
SU(2)$_{W_L}\times$SU(2)$_{W_R}$ symmetry 
into the standard model symmetry SU(2)$_{W_L}\times$U(1)$_Y$ is 
closely connected with the production and decay of hyperquarks
and thus to some extent explained in the present model.

At the grand unification scale $M_{G}\cong 10^{16}$ GeV the spectrum 
of bosonic preon bound states is much larger. 
In addition to the usual GUT bosons, which provide transitions between 
quarks and leptons as in proton decay processes, several new gauge bosons 
appear, the counting of which leads to the grand unification group SU(9)$_{G}$.
Next to the 35 gauge bosons of SU(6)$_{P}$, 
there are 8 gluons of SU(3)$_{C}$, 
18 colored bosons ($X$, $Y$, and $U$ and their antiparticles),
18 hypercolored bosons ($\tilde{X}$, $\tilde{Y}$, and $\tilde{U}$ 
and their antiparticles), and the photon $A_{Q}$ of U(1)$_Q$, altogether 80
generators of SU(9)$_{G}$.

The hyperquark mass scale has been found 
from the dimension of these unified gauge groups, the number of 
fermionic bound states,
and the requirement that the coupling constants of the various 
effective gauge interactions are equal at the two unification scales. 
These constraints lead to a system of three equations with three unknows,
allowing the determination of both unification scales 
$M_{P}\cong 10^{9}$ GeV and 
$M_{G}\cong 10^{16}$ GeV, and the average 
hyperquark mass scale $m_{hq}\cong 10^{4}$ GeV. To obtain a mass constraint 
for the lightest hyperquark bound state, the hyperpion, 
we have extrapolated from the point 
$m_{hq}$ to the point $\Lambda_{H}$ where 
$\alpha_{H} \cong {\cal O }(1)$ and obtained $\Lambda_{H} \cong 1700$ GeV,
as the typical scale for hyperhadron masses, 
which is within reach of the LHC at CERN. 

Finally, from the unification constraints mentioned above, 
the Majorana description of neutrinos is prefered.
For Dirac neutrinos the present theory leads to a proton lifetime 
which is too short, and furthermore to a hyperquark mass scale 
below the masses of W and Z bosons, both results contradicting 
experimental facts. 

In summary, based on the introduction of hyperquarks as a new class
of fermionic bound states we have depicted a scenario for the 
unification of forces that partly explains the left-right asymmetry of weak 
interactions at low energies and fills the large gap between the Fermi scale 
and partial unification scale with a wide spectrum of hyperquark bound 
states. 

\vspace{0.3 cm}

\noindent
{\bf Acknowledgments}: We thank Haim Harari and Don Lichtenberg 
for helpful comments.

\end{document}